\pdfoutput=1
\documentclass[PAPER, cernpreprint, atlasdraft=false, UKenglish, texmf, orcidlogo]{atlasdoc}
\usepackage[subcaption=true]{atlaspackage}
\usepackage{atlasbiblatex}

\usepackage[jetetmiss=true]{atlasphysics}

\usepackage{multirow}

\newenvironment{DIFnomarkup}{}{}

\graphicspath{{logos/}{figures/}}

\usepackage{ANA-EXOT-2021-33-PAPER-defs}

\AtlasTitle{Search for high-mass resonances in photon-jet final states using 140 fb$^{-1}$ of $pp$ collisions at $\sqrt{s} = 13$ TeV with the ATLAS detector}

\AtlasAbstract{%
A search for new high-mass phenomena in events with a high-transverse-momentum photon and a jet is performed using data collected by the ATLAS experiment from proton--proton collisions at the LHC, with a centre-of-mass energy of 13 TeV. The analysis is based on a total integrated luminosity of 140 fb\(^{-1}\). The $\gamma$--jet invariant mass spectrum measured in data is examined for localised excesses above the Standard Model background. No significant deviation from the Standard Model prediction is observed, allowing stringent limits to be set on the production cross sections of specific new physics scenarios.
To investigate quark compositeness, excited quark models are studied for different couplings ($f$) and flavours: light $q_{\mathrm{L}}^{*}$ (\(u^{*}/d^{*}\)) and heavy ($b^{*}$, $c^{*}$). For $f=1$, masses up to 6.0 TeV ($q_{\mathrm{L}}^{*}$), 3.5 TeV ($c^{*}$), and 2.6 TeV ($b^{*}$) are excluded.
These results provide the first direct constraints on $c^{*}$ production in the $\gamma$--jet final state at the LHC.
Additionally, the analysis excludes quantum black hole mass thresholds up to 5.3 (7.5) TeV in the Randall-Sundrum (Arkani-Hamed-Dimopoulos-Dvali) models at a 95\% confidence level, further constraining these theoretical frameworks.
Constraints on the visible cross-section are also obtained using generic Gaussian-shaped resonance production.%
}

\AtlasRefCode{EXOT-2021-33}

\PreprintIdNumber{CERN-EP-2026-159}

\AtlasJournal{\JHEP}
\AtlasCoverEgroupAnalysisTeam{atlas-exot-2021-33-analysis-team@cern.ch}

\hypersetup{pdftitle={ATLAS document},pdfauthor={The ATLAS Collaboration}}

\begin{document}

\maketitle

\tableofcontents

\section{Introduction}
%


Final states with a photon and a jet at large invariant mass (\myj) may be produced in proton--proton (\pp) collisions at the Large Hadron Collider (LHC) in many scenarios of physics beyond the Standard Model (SM). These have been proposed to tackle unresolved questions of the SM, including the limitations to only three fermion families and the
hierarchy problem.%
For example, models based on quark and lepton compositeness have been proposed to explain the existence of the three fermion families, which naturally introduce the existence of excited states~\cite{PhysRevD42815,10.1007/978-3-7091-8830-9_8,Baur:1987ga}. Excited quark states (\qstar) with vector-like couplings to SM particles can be produced in \pp collisions via the fusion of a gluon with a quark followed by a prompt decay into a quark and a photon ($qg \to \qstar \to q \gamma$)~\cite{Bhattacharya:2009p2168, PhysRevD42815}, leading to a resonant peak at the \qstar mass in the \myj distribution. This model serves as a benchmark for the search presented in this paper, considering states of light-flavour \lstar (\(u^{*}/d^{*}\)) and heavy-flavour \bstar and \cstar quarks.
The model is characterised by the excited quark mass \mq, after setting the compositeness scale $\Lambda =  \mq$, and form-factors owing to composite dynamics for SU(3), SU(2) and U(1) couplings $f_s = f = f'$, where $f=1$ corresponds to SM-like couplings strength.

Theories involving extra spatial dimensions, such as the Randall-Sundrum type-I (RS1) model~\cite{Meade:2007sz} and the Arkani-Hamed-Dimopoulos-Dvali (ADD) model~\cite{ArkaniHamed:1998rs}, propose solutions to the mass hierarchy problem by lowering the fundamental scale of quantum gravity $M_{*}$ in 4 + $n$ dimensions ($n$ being the number of extra spatial dimensions) to a few TeV. Consequently, high-energy $pp$ collisions could produce microscopic black holes with masses close to $M_{*}$~\cite{Dimopoulos:2001hw,Giddings:2001bu}. They can be quantum black holes (QBHs)~\cite{Meade:2007sz,Calmet:2008dg} that decay before thermalizing, yielding low-multiplicity final states~\cite{Gingrich:2009hj}. In this case, a broad resonance-like structure would be observed in the {\myj} distribution above the threshold mass $M_\mathrm{th}$ on top of the SM background. Several model parameter choices are possible.
The number of extra spatial dimensions $n$ was set to $n=6$ for the ADD model and to $n = 1$ for the RS1 model. The black hole production threshold $M_\mathrm{th}$  was set equal to the Planck scale $M_{*}
$, with the maximum black hole mass capped at either 3 $M_{*}$ or the LHC {\pp} centre-of-mass energy of \(13 ~\tev\), whichever is lower.  This limit ensures that the black hole mass remains near the Planck scale, avoiding the thermal regime where the classical black hole description and its decay into high-multiplicity final states would apply.

Additionally, the search is interpreted with an approach based on generic Gaussian-shaped resonance mass distribution
to provide information for reinterpretation in other models. Resonance widths ranging from the detector {\myj} resolution of about \(2\%\) to \(15\%\) are explored.

A previous search for excited quarks and QBHs in the {\yj} final state was performed by the ATLAS Collaboration using a \(36.7~\ifb\) dataset of \pp collisions at \(\sqs=13~\tev\)~\cite{EXOT-2016-26}, where the mass range below 5.3 TeV for excited quarks and 7.1 TeV (4.4 TeV) for QBH production in the ADD (RS1) is excluded.
Limits on models with excited quark and QBH production in the {\yj} final state were obtained by the CMS Collaboration using \(138~\ifb\) at \(\sqs=13~\tev\) ~\cite{CMS-EXO-20-012}.
In a recent study ~\cite{CMS-EXO-24-028}, CMS further presented limits on models with extra spatial dimensions, expressed in terms of the fundamental Planck scale $M_D$ and the minimum black hole mass for various choices of model parameters.
Dijet resonance searches conducted by ATLAS~\cite{EXOT-2016-21,EXOT-2016-33,EXOT-2019-03} and CMS~\cite{CMS-EXO-15-001,CMS-EXO-19-012,CMS-EXO-20-008,CMS-EXO-24-011} also set limits on the production cross-sections of excited quarks and quantum black holes.

The search presented in this  paper is  based on the full Run 2 data set recorded with the ATLAS detector  corresponding to  an integrated luminosity of \(140~\ifb\) of \pp collisions at \(\sqs=13~\tev\).
The analysis looks for a resonant excess in the invariant mass of the \yj system over the  SM background. The sensitivity is enhanced compared to the previous search using \(36.7~\ifb\) due to the increased size of the dataset, new criteria in the event selection, and an improved background modelling technique. In addition, the search for excited quarks is performed considering states of light and heavy-flavour quarks and the dependence of the cross section $\times$ branching ratio on the coupling $f$ is explored. The results for \cstar signals are the first ones in the \yj final state at the LHC.

The paper is organized as follows. \Sect{\ref{sec:detector}} briefly describes the ATLAS detector. \Sect{\ref{sec:datasets}} summarises the data and simulation samples used in this study. The event reconstruction and selection criteria are discussed in \Sect{\ref{sec:selection}}. Signal search and limit-setting strategies are detailed in \Sect{\ref{sec:statistical_procedure}}, including the modelling of signal and background. The systematic uncertainties are detailed in \Sect{\ref{sec:stat_sys}}.
The results and their interpretation are presented in \Sect{\ref{sec:results}}, followed by the conclusions in \Sect{\ref{sec:conclusion}}.


\section{ATLAS detector}
\label{sec:detector}

%
%
%
%
%
%
%
%
%
%
%

%
%
%
%

%
%
%
%
%
%
%
%
%
%

%
%
%
%

%
%
%
%
%
%

%
%
%

%
\newcommand{\AtlasCoordFootnote}{%
ATLAS uses a right-handed coordinate system with its origin at the nominal interaction point (IP)
in the centre of the detector and the \(z\)-axis along the beam pipe.
The \(x\)-axis points from the IP to the centre of the LHC ring,
and the \(y\)-axis points upwards.
Polar coordinates \((r,\phi)\) are used in the transverse plane,
\(\phi\) being the azimuthal angle around the \(z\)-axis.
The pseudorapidity is defined in terms of the polar angle \(\theta\) as \(\eta = -\ln \tan(\theta/2)\) and is equal to the rapidity
$ y = \frac{1}{2} \ln \left( \frac{E + p_z}{E - p_z} \right) $ in the relativistic limit.
Angular distance is measured in units of \(\Delta R \equiv \sqrt{(\Delta y)^{2} + (\Delta\phi)^{2}}\).}

The ATLAS experiment~\cite{PERF-2007-01} at the LHC is a multipurpose particle detector
with a forward--backward symmetric cylindrical geometry and a near \(4\pi\) coverage in
solid angle.\footnote{\AtlasCoordFootnote}
It consists of an inner tracking detector surrounded by a thin superconducting solenoid
providing a \qty{2}{\tesla} axial magnetic field, electromagnetic and hadronic calorimeters, and a muon spectrometer.
The inner tracking detector covers the pseudorapidity range \(|\eta| < 2.5\).
It consists of silicon pixel, silicon microstrip, and transition radiation tracking detectors.
Lead/liquid-argon (LAr) sampling calorimeters provide electromagnetic (EM) energy measurements
with high granularity within the region \(|\eta|< 3.2\).
A steel/scintillator-tile hadronic calorimeter covers the central pseudorapidity range (\(|\eta| < 1.7\)).
The endcap and forward regions are instrumented with LAr calorimeters
for EM and hadronic energy measurements up to \(|\eta| = 4.9\).
The muon spectrometer surrounds the calorimeters and is based on
three large superconducting air-core toroidal magnets with eight coils each.
The field integral of the toroids ranges between \num{2.0} and \qty{6.0}{\tesla\metre}
across most of the detector.
The muon spectrometer includes a system of precision tracking chambers up to \(|\eta| = 2.7\) and fast detectors for triggering up to \(|\eta| = 2.4\).
The luminosity is measured mainly by the LUCID-2~\cite{LUCID2} detector, which is located close to the beampipe.
A two-level trigger system is used to select events~\cite{TRIG-2016-01}.
The first-level trigger is implemented in hardware and uses a subset of the detector information
to accept events at a rate close to \qty{100}{\kHz}.
This is followed by a software-based trigger that
reduces the accepted rate of complete events to \qty{1.25}{\kHz} on average,
depending on the data-taking conditions.
A software suite~\cite{SOFT-2022-02} is used in data simulation, in the reconstruction
and analysis of real and simulated data, in detector operations, and the trigger and data acquisition
systems of the experiment.


\section{Dataset and simulated events}
\label{sec:datasets}

The analysis is performed on a set of proton--proton collision data collected
by the ATLAS detector at $\sqs=13~\tev$ between 2015 and 2018.
With requirements on the stability of the beams, the operational status of all
ATLAS detector components, and the quality of the recorded data~\cite{DAPR-2018-01}, the total integrated
luminosity of the dataset is \(140~\ifb\).
The uncertainty in the combined 2015--2018 integrated luminosity is 0.83\% ~\cite{DAPR-2021-01}, obtained using the LUCID-2 detector~\cite{LUCID2} for the primary luminosity measurements, complemented by measurements using the inner detector and calorimeters.

Monte Carlo samples of simulated events are used to study the background modelling for the dominant {\yj} processes, to optimise the selection criteria, and to evaluate the acceptance
and selection efficiencies for the signals considered in the search.
Samples of the expected excited quark, QBH signals, and SM backgrounds were simulated using
dedicated MC generators. The excited quark and QBH signal samples were processed with a fast simulation of the ATLAS detector (AtlFast3)~\cite{SIMU-2018-04}. In contrast, the SM background samples were processed with the {\GEANT}-based full simulation~\cite{Agostinelli:2002hh, SOFT-2010-01}. All the samples were reconstructed with the same algorithms used for the data. An event-by-event weighting is applied to all MC samples to model the realistic experimental conditions of the data sample under study, by matching the simulated distribution of the number of inelastic $pp$ collisions per bunch crossing (pile-up) to the one observed in data. The simulated samples are further corrected with efficiency scale factors as described in \Sect{\ref{sec:selection}}.

The prompt single-photon production was simulated using {\pythia} 8.186~\cite{Sjostrand:2007gs}. The partonic processes are simulated using leading-order (LO) matrix elements, including the initial- and final-state parton showers. Fragmentation into hadrons was performed using the Lund string model~\cite{ANDERSSON198331}. The NNPDF2.3 LO~\cite{Ball:2012cx} for {\Pythia} parton distribution functions (PDFs) were used to parameterise the proton structure. The sample also includes a simulation of the underlying event. The event generator parameters are set according to the A14 tuned set of parameters (tune) for {\Pythia}~\cite{ATL-PHYS-PUB-2014-021}. The {\Pythia} simulation includes leading-order {\yj} diagrams from both direct processes (the hard sub-processes $qg \to q\gamma$ and $qq \to g\gamma$, called ``direct'' component) and photon bremsstrahlung in QCD dĳet events (called ``fragmentation'' component). In the case of the fragmentation component, photons can be radiated in the parton shower without a restriction on the opening angle with respect to the parent parton, and, as a result, the photons can be emitted very close to the parton direction.

Samples of excited quarks for {\lstar}, {\bstar} and {\cstar} at different invariant masses are generated with {\Pythia} 8.245~\cite{Sjostrand:2014zea}. NNPDF2.3 LO parton distribution functions and the A14 tune of the underlying event are used.
Although excited quarks decay predominantly into quark--gluon final states, in the present study only quark--photon final states are generated. The branching ratio into a quark--photon final state is in the range of 0.5-2\% over the mass range considered~\cite{Cakir:1999nu}.
The compositeness scale $\Lambda$ was taken to be equal to the mass $m_{q^*}$ of the excited quark, and the coefficients $f_s$, $f$ and $f'$ of magnetic-transition type couplings to the respective SU(3), SU(2) and U(1) gauge bosons were chosen to be equal; values of 0.01, 0.1, 0.5, 0.75 and 1.0 are simulated.
The {\lstar} samples were generated with $m_{u^*} = m_{d^*}$ values between 0.5 and 7 TeV in steps of 0.5 or 1 TeV. The {\bstar} and {\cstar} samples were generated with mass values between 0.5 and 4 TeV.

Samples of QBH decaying into a photon and a parton were generated with different threshold masses with the QBH version 3.02 described in Ref.~\cite{Gingrich:2009da} and {\Pythia} 8.306~\cite{Bierlich:2022pfr} for hadronisation and underlying event. The CTEQ6L1 PDF set~\cite{Pumplin:2002vw} was used together with the A14 tune.


\section{Event reconstruction and selection}
\label{sec:selection}


%
%

Primary vertices are built from sets of two or more tracks, each with a transverse momentum of \(\pt > 500 ~\mev\). The vertex with the highest value of \(\sum \pt^2\) for all the associated tracks is denoted as the primary vertex~\cite{IDTR-2021-01}.

Photons and electrons are reconstructed using dynamic, variable-sized clusters of energy deposits in the electromagnetic and hadronic calorimeters, topologically connected~\cite{PERF-2014-07,ATL-PHYS-PUB-2017-022}, called \topos.
These are matched to charged-particles tracks in the inner detector.
{\Topos} that are not matched to a track or a conversion vertex are classified as unconverted photons. On the other hand, if a {\topo} is matched to a conversion vertex, it is classified as a converted photon. Finally, when a {\topo} is matched to a track, it is classified as an electron~\cite{EGAM-2018-01}.

Photon candidates are required to pass Loose or Tight identification criteria, based on selections on observables describing the shape of the shower in the EM and hadronic calorimeters, called shower shapes~\cite{EGAM-2018-01}. Baseline photons must pass Loose criteria, have \(\pt > 25~\gev\), and pseudorapidity in the range \(\abseta < 2.37\), avoiding the barrel-endcap transition region (\(1.37 < \abseta < 1.52\)). Signal photons are further required to pass Tight identification criteria and to have a minimum transverse momentum of \(\pt > 150~\GeV\) in order to guarantee the trigger selects the photons.
To further reduce the number of jets that can be misidentified as photons (fake photons), two types of isolation selections are applied to the photon candidates. The first isolation criterion is based on the sum of energies in the calorimeter cells in a cone of size \(\Delta R = 0.4\) around the cluster barycentre, excluding the energy associated with the photon cluster, defining the calorimetric isolation energy \etiso. This \etiso is required to be less than \(0.022\, \pt + 2.45~\gev\).
The second isolation criterion is based on the track isolation variable \ptiso, which corresponds to the scalar sum of the transverse momenta of non-conversion tracks in a cone of size \(\Delta R = 0.2\) around the cluster barycentre, and it is required to be $\ptiso / \pt^{\gamma} < 0.05$.

Electron candidates are required to have \(\pt>10~\gev\) and \(\abseta < 2.47\), and to originate from the primary vertex. They also must pass the Tight identification criterion, based on the EM and hadronic shower shape, the quality of the associated track, and the angular proximity of the track to the calorimeter energy deposition.
The associated track is required to have a longitudinal impact parameter  \(|z_0\sin\theta | < 0.5~\text{mm}\) and a transverse impact parameter significance \(|d_0|/\sigma_{d_0} < 5.0\).
Furthermore, electrons are required to pass a similar calorimetric and track isolation criterion. If any electron remains after the selection, the event is discarded.

Muon candidates are reconstructed from the tracks in the Muon Spectrometer and the inner detector. The candidates are required to originate in the primary vertex, to have \(\pt > 10~\gev\) and \(\abseta < 2.7\), and to pass the Medium identification criterion~\cite{MUON-2018-03}, and impact parameter selections of \(|z_0\sin\theta | < 0.5~\text{mm}\) and \(|d_0|/\sigma_{d_0} < 3.0\). Selected muons must also pass the Loose isolation criterion, which is defined similarly to the electron isolation criterion.
The event is removed if there is a muon after the mentioned selection.

Jets are reconstructed using a particle flow algorithm~\cite{Cacciari:2008gp, Fastjet} exploiting both the ID and calorimeter information. The \antikt algorithm with a radius parameter of \(R = 0.4\) is used to cluster the jets~\cite{PERF-2015-09}. Jets are required to have \(\pt > 20~\gev\) and to be in the pseudorapidity range of \(\abseta < 2.8\). To reject jets originating from pile-up interactions, jets with \(
\pt < 60 ~\gev\) are required to satisfy a neural network jet vertex tagger discriminant, successor of the jet vertex tagger algorithm~\cite{PERF-2014-03}, using track information from the ID.

Heavy flavour jets are particularly relevant in this search, as the flavour of the leading jet is used to specify the signal regions.
The GN2 tagger, based on a transformer~\cite{FTAG-2023-05}, is used to discriminate between light-, \(c\)-, \(b\)- and \tjets. It makes use of the long lifetime, high mass, high decay multiplicity and characteristic decay modes of the \(b\)- and \(c\)-hadrons as well as properties of heavy-quark fragmentation.
Flavour-tagging is applied only to the leading jet and if it satisfies \(\abseta < 2.5\). The algorithm's outputs are the probabilities of the jet being a \ljet, \cjet, \bjet or \tjet. These probabilities are combined into a single discriminant \(D_b\) (\(D_c\)), allowing the definition of different \btagging (\ctagging) efficiency working points.

Since an energy deposit or charged-particle track can contribute to the reconstruction of more than one object, a procedure is applied to baseline objects before isolation requirements to remove possible double-counting: if a muon shares an ID track with an electron, it is removed; if there is a photon within \(\Delta R < 0.4\) of an electron or muon, then the photon is removed; any jet within \(\Delta R < 0.2\) of an electron is removed and any electron within \(\Delta R < 0.4\) of a jet is removed; then, any jet found with an angular distance of \(\Delta R < 0.4\) to a muon with less than three tracks is removed; if there is a muon within \(\Delta R < 0.2\) of a jet, the muon is removed. Finally, jets with an angular distance of \(\Delta R < 0.4\) from a photon candidate are discarded.

The data sample for this analysis is selected using a trigger that requires at least one loose photon with \(\pt > 140~\gev\)~\cite{TRIG-2018-05}. To optimise the signal-to-background ratio, selected events are required to contain at least one signal photon in the central region of the detector (\(\abseta < 1.37\)). The combined photon identification and isolation efficiency for signal events is \(78\%\). Differences between data and MC efficiencies are corrected using data-to-simulation scale factors, which are determined as a function of the photon \pt and \abseta~\cite{EGAM-2018-01}. The selected  events are also required to have at least one jet with  \(\pt > 150~\gev\) and \(\abseta < 1.37\). The leading photon and leading jet are used as the {\yj} pair in the following. Since direct photon production is characterised by the photon and jet carrying approximately the same momentum, the requirement $(\ptjet - \pt^{\gamma}) / \pt^{\gamma} < 0.5$ rejects the bremsstrahlung component.
The invariant mass of the {\yj} system is required to be $m_{\gamma j} > 500~\gev$. To suppress the direct photon  \(t\)-channel production, the pseudorapidity differences between the photon and the jet is restricted to $\detayj = |\eta^{\gamma} - \etajet| <1.6$. Any event with a jet within \(\dryj < 1.0\) is removed to extend the photon's isolation requirements.

This event selection defines the inclusive signal region SR. Furthermore, this search benefits from the  GN2 tagger, allowing one to discern between light-, \(c\)- and \bjets. The \btagging signal region (SRB) is defined  by requiring the leading jet to be tagged as a \bjet using the \(77\%\) efficiency working point. Events in which the jet does not satisfy the SRB criteria, are further classified into the \cjet signal region (SRC) or the {\ljet} signal region (SRL). The SRC includes events where the jet is tagged as \cjet using the \(50\%\) efficient working point, while SRL consists of events where the jet fails the \btagging and \ctagging requirements.
The \(77\%\) \btagging and \(50\%\) \ctagging working points are defined to correspond to the quoted {\bjet} and {\cjet} tagging efficiencies in simulated \ttbar events, however, both efficiencies decrease at higher \pt. When measured in high-\pt jets from simulated $Z'$ samples, the \(77\%\) working point corresponds to a {\bjet} tagging efficiency of $\sim41\%$, with rejection factors (defined as the inverse of the misidentification efficiency) of $\sim20$ and $\sim180$ against {\cjets} and light-flavour jets, respectively. Similarly, the \(50\%\) working point corresponds to a {\cjet} tagging efficiency of $\sim34\%$, with rejection factors of $\sim5$ and $\sim13$ against \bjets and light-flavour jets, respectively~\cite{FTAG-2023-05}.

The complete event selection is shown in \Tab{\ref{tab:evt_selection:srs}}.

\begin{table}[ht!]
\caption{Definition of common selection required for the inclusive signal region SR.
The three exclusive signal regions SRB, SRC and SRL are distinguished by additional flavour-tagging requirements:
SRB requires the jet to pass the 77\% {\btag} working point and no {\ctag} requirement, SRC requires the jet to pass the 50\% {\ctag} working point and fail the 77\% {\btag} requirement, while SRL requires the jet to fail both the 77\% {\btag} and 50\% {\ctag} requirements.}
\label{tab:evt_selection:srs}
\centering
\begin{tabular}{r>{\centering\arraybackslash}p{0.08\linewidth}|>{\centering\arraybackslash}p{0.08\linewidth}|>{\centering\arraybackslash}p{0.08\linewidth}}

\toprule
\multicolumn{4}{c}{\textbf{SR}} \\
\midrule

Photon: & \multicolumn{3}{l}{$\pt^{\gamma} >150$ GeV, $|\eta^{\gamma}|<1.37$} \\
Jet:    & \multicolumn{3}{l}{$\ptjet >150$ GeV, $|\etajet|<1.37$} \\

\multicolumn{4}{c}{$m_{\gamma j} > 500\, \gev$} \\
\multicolumn{4}{c}{$\detayj                                 <1.6$} \\
\multicolumn{4}{c}{$\Delta R_{\text{min}} (\gamma, \text{jet}) > 1.0$} \\
\multicolumn{4}{c}{$(\ptjet - \ptgam)/\ptgam <0.5$} \\
\midrule
\multicolumn{1}{c|}{}                              &  \textbf{SRB}   & \textbf{SRC} & \textbf{SRL} \\
\midrule
\multicolumn{1}{c|}{\btag \(77\%\)}                &  $\checkmark$   & $\times$ & $\times$ \\
\multicolumn{1}{c|}{\ctag \(50\%\)}                &       -         & $\checkmark$ & $\times$ \\
\bottomrule
\end{tabular}
\end{table}


\section{Statistical analysis}
\label{sec:statistical_procedure}

Fits to the {\myj} spectra are performed using a binned likelihood method. The likelihood function {\LL} is built from a Poisson probability distribution describing the number of observed events in each bin of the selected sample and incorporating systematic uncertainties with Gaussian or log-normal constraint functions.
The fit includes expected signal from MC simulation and background counts estimated from a functional fit to
data.
The signal strength parameter \(\mu\) acts as a scaling factor for the signal yield across all bins. For a parametrised background, the expected background count in each bin can be expressed as a function of the observable \myj, with the total normalisation and shape parameters serving as nuisance parameters.
The number of signal events in each bin is determined by the product of the signal probability density function,  the cross-section times branching ratio, %
the total integrated luminosity, and the signal acceptance times efficiency.  In the absence of a signal \(\mu = 0\), the fit corresponds to a background-only hypothesis, providing a baseline for comparison.

The uncertainty associated with the choice of the functional model for the background is accounted for by the so-called ``spurious signal''  ($S_{\text{spur}}$),  which arises from a non-zero extracted signal when fitting a dataset with no true signal~\cite{HIGG-2013-08}. %
To incorporate this effect, an additional term ($S_{\text{spur}} \theta_{\text{spur}}$) is introduced in the total event count $\mu s \to \mu s + S_{\text{spur}} \theta_{\text{spur}}$
where $\theta_{\text{spur}} $ is a nuisance parameter associated with the background modelling uncertainty and $s$ is the number of signal events.

A significant advancement in this analysis, compared to previous iterations, is the improved capability to distinguish between jets initiated by light quarks (\(u\), \(d\)) and those initiated by heavy quarks (\(c\), \(b\)).
While the separation of \bjets from other jet types is highly effective, distinguishing \cjets remains more challenging. As a result, a \cstar signal would populate not just SRC, but also SRL and SRB. To better exploit this information, a simultaneous fit approach across multiple signal regions is pursued to improve the search sensitivity for excited \cstar states. The combined likelihood function for all signal regions is constructed as the product of the individual likelihoods for each category, ensuring a comprehensive statistical interpretation of the results.  The experimental systematic uncertainties that affect only the signal are fully correlated across all signal regions. In contrast, the spurious signal uncertainty is specific to each signal region, as it depends on the choice of background function and fit range used in the analysis.
For the remaining searches, the fit of the \myj spectra is performed separately for each signal region: the inclusive SR (SRB) is used for the search of \lstar (\bstar); the SR is also used for the search of QBH. For generic Gaussian signal shapes, all signal regions are considered independently.

%
%
%
%
%
%
%
%
%
%
%
%
%
%
%
%
%
%
%
%
%
%
%
%
%
%
%
%
%
%
%
%
%
%
%
%
%
%
%
%
%
%
%
%
%


\subsection{Signal modelling}
%

The results are interpreted using model-independent and model-dependent strategies. The invariant mass distribution \myj for the signal is expected to peak near the assumed mass of the new particle, with a spread given by the convolution of its intrinsic decay width with the
experimental resolution. The experimental resolution of the invariant mass is modelled with a Gaussian function. Interference effects between signal and background are ignored. For the model-dependent limits, the \myj probability density function (pdf) for both excited quark states and QBHs are built starting from the reconstructed MC events passing the selection requirements described in \Sect{\ref{sec:selection}} and applying a kernel density estimation ~\cite{TMVA} to smooth the distributions. Once all the pdfs are obtained, signal models for intermediate mass points  or other couplings are obtained with a moment morphing method as described in \Refn{\cite{Baak:2014fta}}. For the model-independent results, the signal templates are modelled as normalised Gaussian distributions with the mean located at $\mG$. The standard deviation ($\sigma_G$) of the Gaussian distribution is chosen to be  \(2\%\), \(7\%\) or \(15\%\) of \mG, where \(2\%\) corresponds approximately to the effect of the detector resolution on the reconstruction of the {\yj} invariant mass.

The combined acceptance times efficiency ($A \times \varepsilon $) for the case of excited light quarks increases from  29\% at low masses to 43\% for masses above  2 \tev, while for heavy quarks, it ranges from 13\% to 20\% for {\cstar}
and from 17\% to 26\% for {\bstar}, in their corresponding signal region, as illustrated in \Fig{\ref{fig:signal_modelling:Nom__efficiency_acceptance__Run2}}.  For QBH signals, the values range between 34\%-43\% for RS1 model and 40\%-43\% for the ADD model across different QBH mass ranges, also shown in \Fig{\ref{fig:signal_modelling:Nom__efficiency_acceptance__Run2}}.

\begin{figure}[ht!]
\centering
\begin{subfigure}[h]{0.49\linewidth}
\centering
\includegraphics[width=\linewidth]{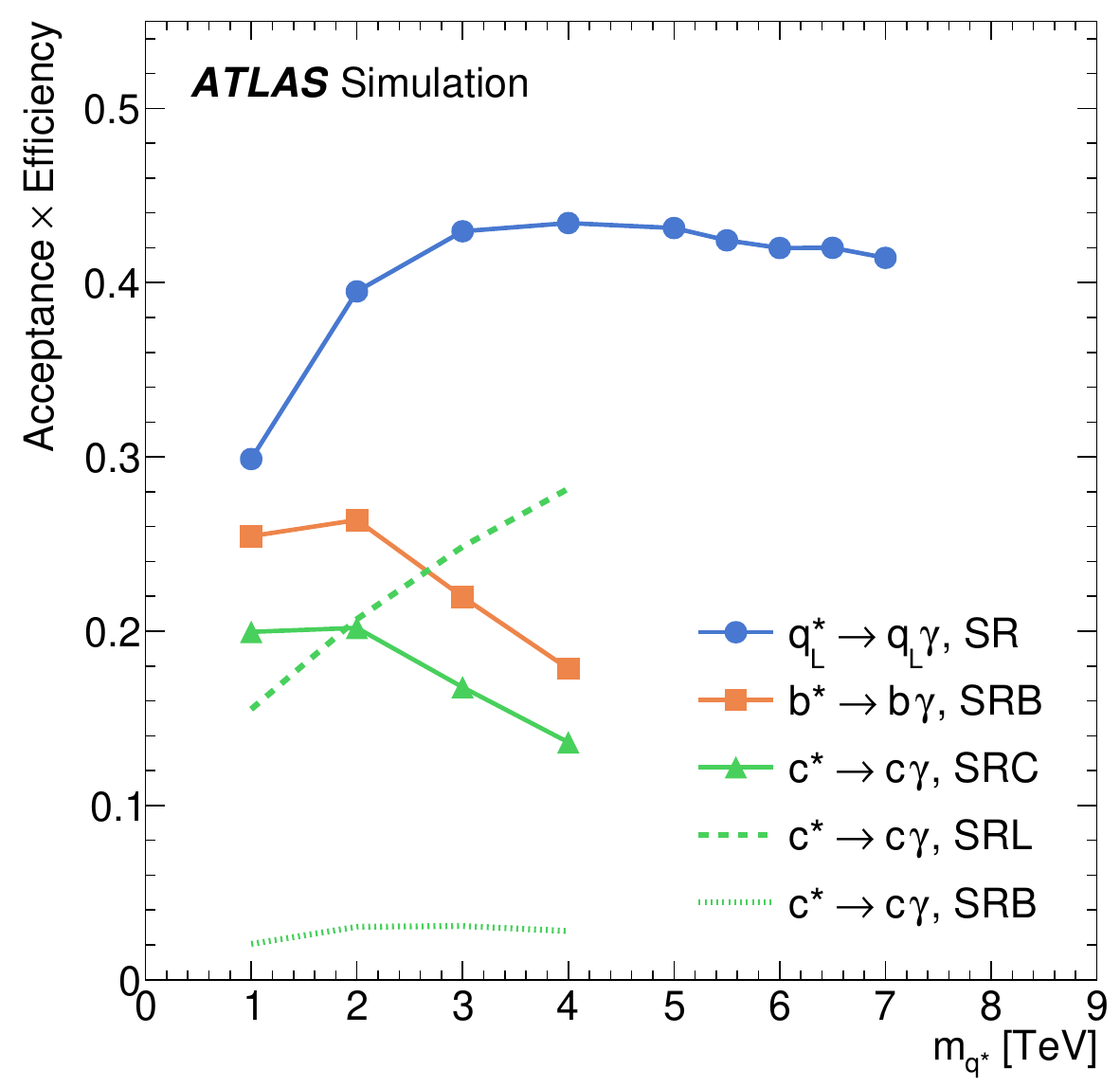}
\caption{}
\end{subfigure}
\begin{subfigure}[h]{0.49\linewidth}
\centering
\includegraphics[width=\linewidth]{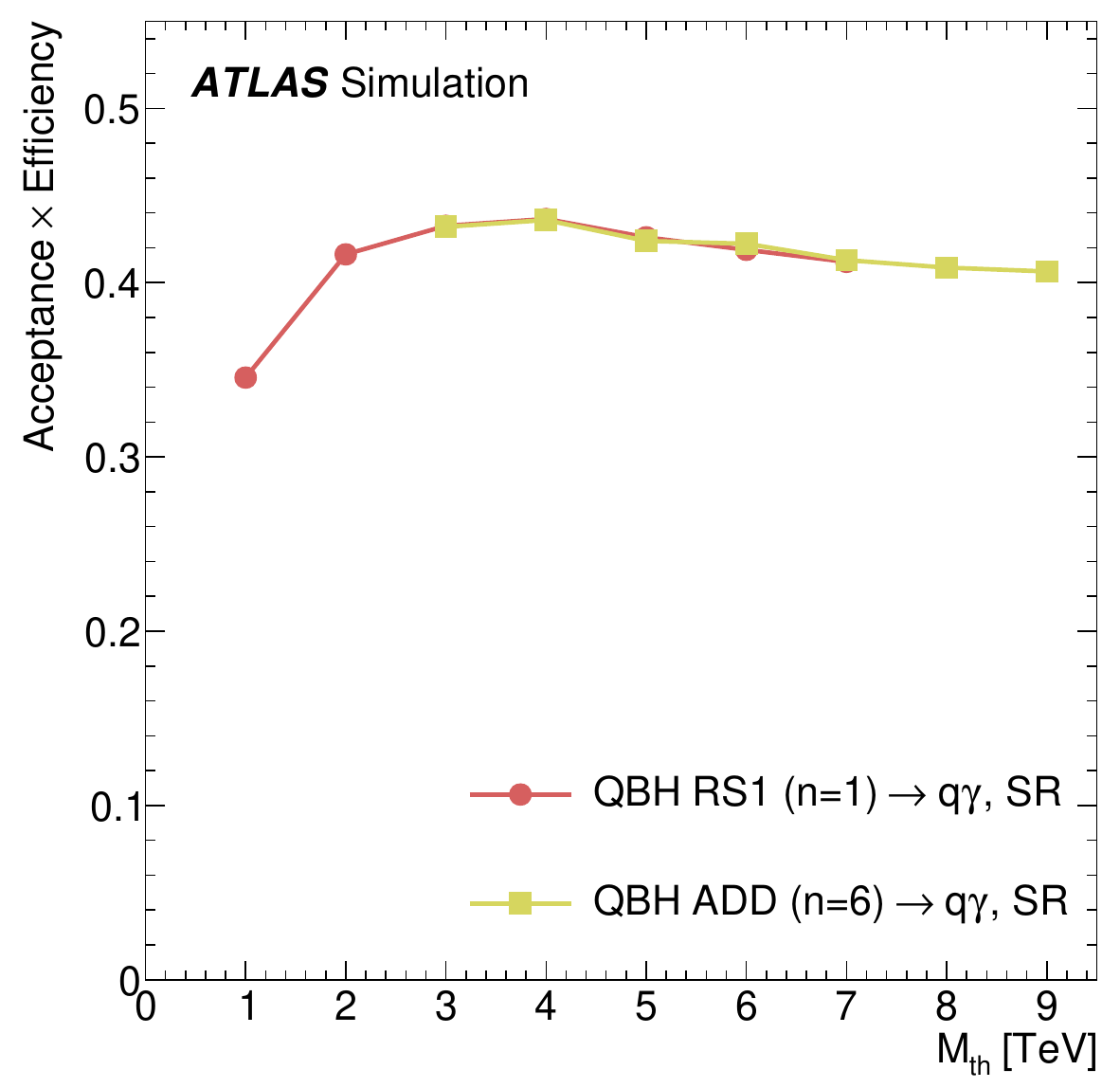}
\caption{}
\end{subfigure}
\caption{Acceptance $\times $ selection efficiencies for the signal models considered in the analysis
as a function of the (a) excited-quark mass or the (b) QBH threshold mass. The description of the selection criteria can be found in the text.}
\label{fig:signal_modelling:Nom__efficiency_acceptance__Run2}
\end{figure}


\subsection{Background modelling}
%


%

The background in the \yj signal extraction fit, consisting of prompt photon and multijet events where one of the jets is misidentified as a photon, is modelled analytically. The functional form chosen for the background model is based on background templates built using simulations and data in control-regions that are also used in the estimation of related systematic uncertainties. The following family of smoothly-falling functions are used~\cite{EXOT-2016-26, EXOT-2022-18}:
\begin{equation}
f_{b}( x \equiv m_ {\gamma j}/ \sqrt{s}) = n_{\text{bkg}\, }(1-x)^{p_0} x^{-\sum_{i=1} p_i \left(\ln x\right)^{i-1}}
\end{equation}

where $\sqrt{s}$ is the centre-of-mass energy, $p_0$ and $p_i$ are the shape parameters, and  $n_{\text{bkg}} $ is a normalisation factor.  The parameters of these functions are simultaneously determined by the final fit of the signal plus background model to data, while the functional forms are chosen from dedicated studies as described in this section.

After applying the photon identification and isolation requirements to the data, a residual reducible background remains, arising primarily from jets misidentified as photons in multijet events.
This contribution is estimated using a data-driven technique based on the two-dimensional sideband method described in Refs. \cite{SUSY-2016-27,STDM-2017-01}, and is found to be an order of magnitude smaller than the \yj background.
The method relies on counting photon candidates in four regions of a two-dimensional plane defined by the  identification criteria (tight or non-tight) and calorimetric isolation (isolated and non-isolated).
The photon signal region (region A) is defined by photon candidates that are isolated and satisfy the tight identification criteria. Three control regions (CRs) are defined accordingly: tight and non-isolated (region B), non-tight and isolated (region C), and non-tight and non-isolated (region D).
Photon candidates are classified as non-tight if they fail the tight identification criteria but satisfy a modified set of requirements based on a subset of the shower-shape variables measured in the first layer of the EM calorimeter. A gap of \(8~\gev\) in the isolation variable is introduced to reduce signal contamination in the CRs. Residual correlations between identification and isolation, as well as contamination from prompt photons in region B, C, and D, are taken into account. This procedure enables the determination of factors used to estimate the contribution of fake photons in all analysis regions.
To determine the yields of real and fake photons in the regions defined by the sideband method, a sequential template fit procedure to the photon isolation distribution is performed in data and simulations.
Templates for prompt photons are derived from simulated {\pythia} samples, while templates for fake photons are obtained from data control samples.
The resulting fake factors are applied to data in a CR that differs from the SR only in the photon isolation requirement.
The corresponding fake-photon distributions are then fitted with a higher-order functional form than the one used in the analysis fit, in order to obtain smooth parametrisations.

Two types of samples are used for background modelling studies: Asimov datasets~\cite{Cowan:2010js} and pseudo-data. The high-statistics \pythia  \yj sample, supplemented with a smoothed jet-faking-photon distribution, is fitted to produce the Asimov datasets. Since multiple combinations of functional models (with varying degrees of freedom) and fit ranges are defined for each signal region, the prepared {\myj} distributions are fitted with different combinations to generate the Asimov datasets. Pseudo-data distributions are then obtained from background-only fits to the Asimov datasets that satisfy $\chi^2$ {\pval} $> 0.05$. For each signal region and functional model, multiple pseudo-data sets are generated according to the number of parameters in the background-only fit. To test fits using a model with $n$ parameters, pseudo-data distributions generated from a model with $n+1$ parameters are used.

The spurious signal test assesses whether the fitting procedure is biased, producing a non-zero extracted signal when fitting a dataset with no true signal~\cite{HIGG-2013-08}. Spurious signals are used in the final limit calculation as the systematic uncertainty associated with the background modelling. This test is performed on either the Asimov dataset or the full set of pseudo-data distributions.
Results from pseudo-data are used up to the mass point where a significant background presence is expected, for regions dominated by {\ljets} ({\bjets} and {\cjets}) the maximum mass point considered is $<5$ \tev ($<4$ \tev), while Asimov dataset are used in the remaining mass range. Each distribution is evaluated by conducting signal+background fits across each signal mass, for each model, and within each signal region. For selection in the analysis, the functional form must exhibit a spurious signal that is less than \(50\%\) of the statistical uncertainty in the fitted signal yield over the entire mass range under investigation. Among functional forms meeting this criterion, the one with the fewest degrees of freedom is preferred.
For different signal models, the functional form and {\myj} fit range are determined independently. Models and mass ranges that pass this test proceed to the signal injection tests.

The signal injection test is performed to confirm that the fit can accurately extract a signal component with the expected strength.  In this test, simulated signal models are added to the fitted background distribution with a selected signal cross-section
ranging from 0 to \(5 \zeta\), where \(\zeta\) is defined as \(S/\sqrt{B}\) and the number of signal ($S$) and background ($B$) events are determined using a \(2 \sigma\) window around the injected signal peak in each test. The definition of the spurious signal tests remains the same; however, in this case, the spurious signal is evaluated relative to the injected signal ($\sspur = S_{\text{fit}} - S_{\text{inj}}$).
These tests were satisfied for all models over the same range of resonance masses derived from the spurious signal tests.

To determine whether a more complex function is necessary to describe the data, an \(F\)-test
is performed. Two background-only fits are carried out on pseudo-data: one using the simplest validated function and another with a more complex version that incorporates additional parameters. These fits are conducted within the ranges established by the previous spurious signal test. The results do not indicate a need for extra degrees of freedom compared to the simplest functions considered across the entire fit range for all models.

A summary of the functional forms and fit ranges provided for each model and signal region is presented in \Tab{\ref{tab:background:functions}}.
Interpretations of the \lstar model are based on the inclusive signal region (SR). For the \bstar model, the highest significance is observed in the SRB region, leading to the decision to use only SRB for its interpretation. Finally, as discussed in \Sect{\ref{sec:statistical_procedure}}, an improvement in significance for \cstar signals is achieved by simultaneously fitting in the three regions: SRL, SRB, and SRC.

\begin{table}[ht!]
\caption{Summary of the fit functions and mass ranges (in \gev) used for each signal model and analysis region.}
\label{tab:background:functions}
\resizebox{\linewidth}{!}{
\begin{tabular}{lcccc}
\toprule
& SR                      & SRL                               & SRC                               & SRB                               \\
\midrule
Background-only                & 2-param, \([1000,8000]\)    & 2-param, \([1200,8000]\)    & 4-param, \([1100,7000]\)    & 2-param, \([1100,6000]\)\\
\midrule
Gaussians                & 3-param, \([1200,8000]\)    & 3-param, \([1500,8000]\)    & 4-param, \([1000,7000]\)    & 5-param, \([900,7000]\)\\
QBH                      & 5-param, \([1000,8000]\)    & -                                 & -                                 & - \\
Excited Quarks, {\lstar}  & 2-param, \([1000,8000]\)    & -                                 & -                                 & -\\
Excited Quarks, {\cstar}  & -                                 & 2-param, \([1200,8000]\)    & 4-param, \([1100,7000]\)    & 2-param, \([1100,6000]\)\\
Excited Quarks, {\bstar}  & -                                 & -                                 & -                                 & 2-param, \([1100,6000]\)\\
\bottomrule
\end{tabular}
}
\end{table}


\section{Systematic uncertainties}
\label{sec:stat_sys}

Experimental uncertainties stem from limitations in detector simulation, object reconstruction and calibration, pile-up modelling and luminosity determination. These uncertainties affect only the signal yield, while the effects on signal shape was determined to be negligible.

The uncertainty on the combined 2015-2018 integrated luminosity is \(0.83\%\)~\cite{DAPR-2021-01}, determined using the LUCID-2 detector~\cite{LUCID2} as the primary tool for luminosity measurements.
Uncertainties arising from the rescaling of MC events to match the pile-up distribution observed in data affects the signal efficiency by around 1.5\%. The systematic uncertainties in the photon identification and isolation efficiencies are estimated following the prescriptions in Ref.~\cite{EGAM-2018-01}. They are evaluated by varying the correction factors for photon selection efficiencies in simulation within their corresponding uncertainties, amounting to less than 0.6\% for identification and 1.3\% for isolation.  The experimental uncertainties in the photon energy scale and resolution have an impact on the selection efficiency remaining below 0.2\% for all models studied.
Systematic uncertainties in the jet energy scale and resolution are assessed using a combination of in situ measurements and simulation-based techniques, as detailed in~\cite{PERF-2014-07,PERF-2015-09,JETM-2018-05}.
Additional systematic uncertainties are considered for final states that impose requirements on the number of $b$($c$)-tagged jets.
The uncertainty in the $b$-tagging ($c$-tagging) efficiency affects the signal event selection efficiency by 3\% (6\%). Systematic variations are considered for the excited quark signal samples using different  PDF families in event generation.
The change in the signal yield  is recalculated for each provided PDF variation, with the standard deviation of all such variations serving as a measure of the systematic uncertainty due to PDFs. Its impact on the selection efficiency of the excited quark signal samples is found to be 0.3\%, 0.5\% and 0.6\% for {\lstar}, {\cstar} and {\bstar}, respectively.
Following previous publications, no theoretical uncertainties are assigned to the QBH benchmark signals.

A summary of the dominant systematic uncertainties in the $A\times\varepsilon$ as described above and included in the statistical analysis is given in \Tab{\ref{tab:stat_unc:su}}.
Since the signal uncertainties are mass-dependent, the listed values represent the maximum across the mass range considered.

\begin{table}
\centering
\caption{Summary of dominant systematic uncertainties in the $A\times\varepsilon$ included in the fit model.}
\label{tab:stat_unc:su}
\begin{tabular}{lccccc}
\toprule
Source                  & {\lstar}    & {\cstar}  & {\bstar}    & QBH \((n=1)\) & QBH \((n=6)\) \\
\midrule
PDFs                    & \(0.3\%\)  & \(0.5\%\)  & \(0.6\%\)  & -           & -           \\
Pile-up                 & \(1.5\%\)  & \(1.4\%\)  & \(1.4\%\)  & \(1.4\%\)  & \(1.4\%\)  \\
Photon identification   & \(0.5\%\)  & \(0.5\%\)  & \(0.5\%\)  & \(0.5\%\)  & \(0.6\%\)  \\
Photon isolation        & \(1.3\%\)  & \(1.3\%\)  & \(1.3\%\)  & \(1.3\%\)  & \(1.3\%\)  \\
Photon energy resolution& \(<0.1\%\) & \(0.1\%\)  & \(<0.1\%\) & \(<0.1\%\)  & \(<0.1\%\)  \\
Photon energy scale     & \(0.2\%\) & \(0.1\%\) & \(0.1\%\) & \(<0.1\%\)  & \(<0.1\%\)  \\
Jet energy resolution   & \(1.2\%\)  & \(1.1\%\)  & \(1.0\%\)  & \(0.9\%\)  & \(0.9\%\)  \\
Jet energy scale        & \(0.6\%\)  & \(0.5\%\)  & \(0.5\%\)  & \(0.5\%\)  & \(0.4\%\)  \\
\(b\)-tagging (SRB)     & -          & \(3\%\)    & \(3\%\)    & -           & -           \\
\(c\)-tagging (SRC)     & -          & \(6\%\)    & -          & -           & -           \\
\(c\)-tagging (SRL)     & -          & \(4.5\%\)    & -          & -           & -          \\
\bottomrule
\end{tabular}
\end{table}

Finally, spurious signals are used in the limit calculation as the systematic uncertainty associated with the background modelling.


\section{Results}
\label{sec:results}

\subsection{Search results}

The results of the background-only fits are shown in \Fig{\ref{fig:results:bkg_only:bkg_only_fits}} for the four considered signal regions: SR, SRL, SRC, and SRB.
Fits are performed in the ranges and with the functional forms discussed in \Tab{\ref{tab:background:functions}},
using the function that gives the lowest spurious signal.
The data are well-described by the 2-parameter fit function in the SR, SRL and SRB regions, yielding $p$-values of \(0.11\), \(0.62\), and \(0.11\), respectively. In the SRC region, the data are fitted with a 4-parameter function, resulting in a \pval\ of \(0.30\).
The events with the highest invariant mass values observed in data reach values of 3.4 TeV, 4.8 TeV and 5.2 TeV in the signal regions SRB, SRC and SRL, respectively.

To investigate potential resonant signals in the \myj distributions across all signal regions, the \bh algorithm~\cite{2011arXiv1101.0390C}, as implemented in pyBumpHunter~\cite{10.21468/SciPostPhysCodeb.15}, is employed. This analysis uses mass bins, with widths determined by the \myj mass resolution, which is obtained through Gaussian fits to the mass response distribution. For each scanned mass window, \bh evaluates the statistical significance of any observed deviations between data and the background prediction. The \bh \pval\ is defined as the lowest observed probability for the data to deviate from the background prediction within a given window due to Poisson fluctuations, as estimated through pseudo-experiments generated from the background model.

The most significant localised excesses identified by the \bh algorithm are highlighted with filled bins in the lower panels of each signal region in \Fig{\ref{fig:results:bkg_only:bkg_only_fits}}. The global significance and the corresponding \bh \pval (significance), denoted as $p_{\text{BH}}$ ($Z_{\text{BH}}$)  are provided for each signal region. A negative significance indicates that the deviation observed in the data is less significant than that observed in the pseudo-data.  Overall, the background modelling is excellent, with only minor discrepancies observed in the fit residuals, primarily due to downward fluctuations of the data in the high-mass regions.

\begin{figure}[ht!]
\centering
\begin{subfigure}[h]{0.49\linewidth}
\centering
\includegraphics[width=\linewidth]{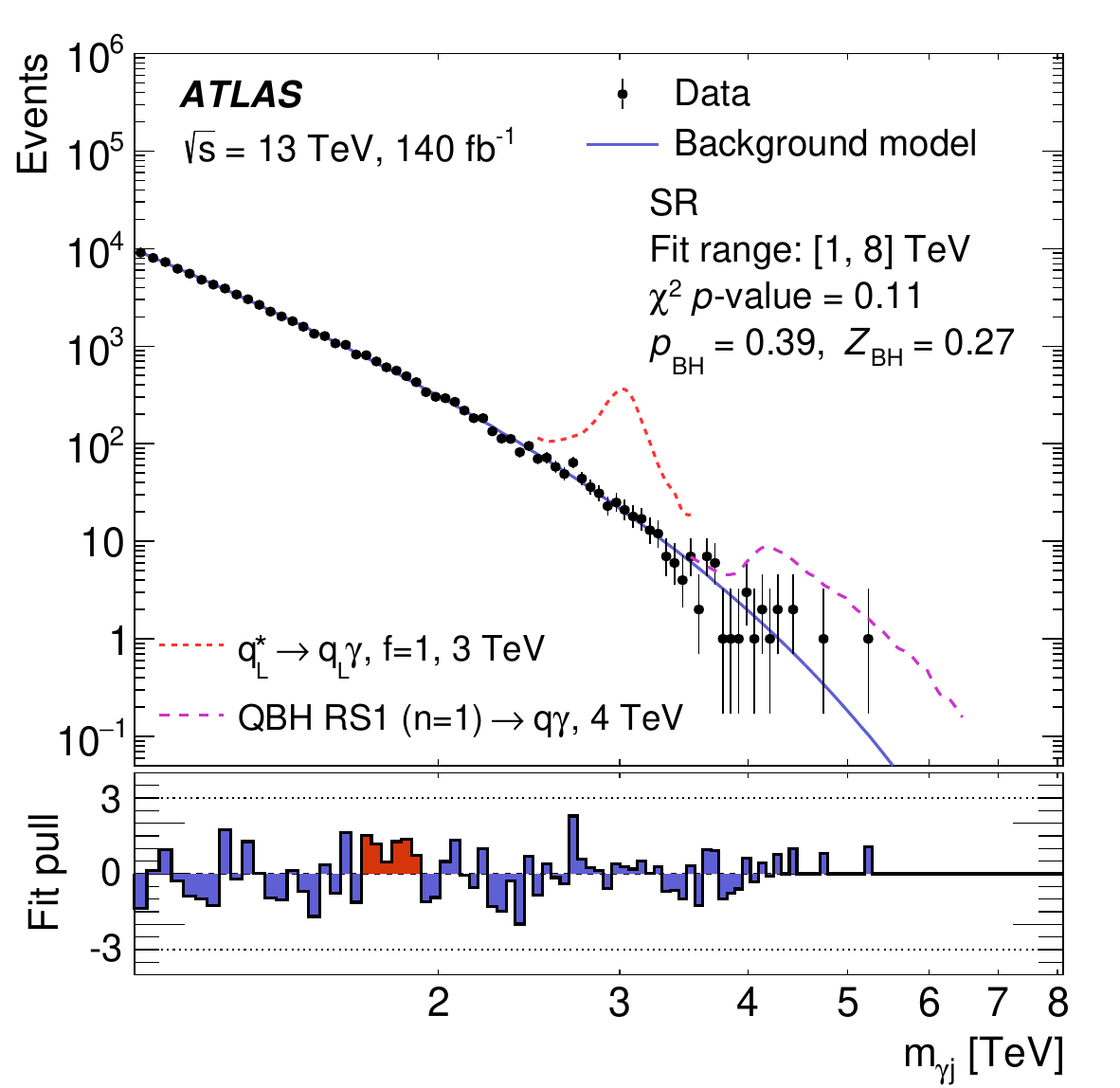}
\caption{}
\end{subfigure}
\begin{subfigure}[h]{0.49\linewidth}
\centering
\includegraphics[width=\linewidth]{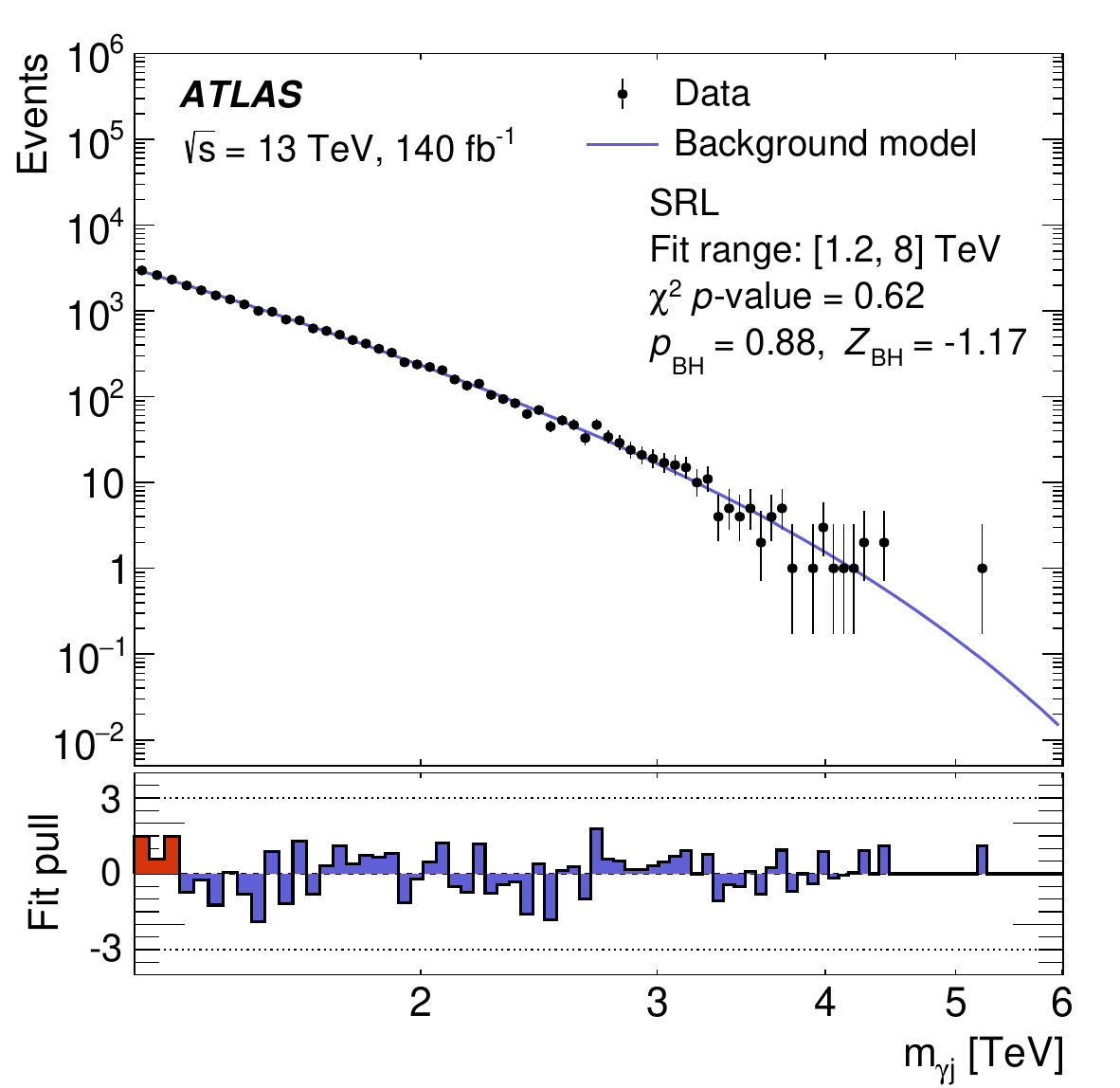}
\caption{}
\end{subfigure}\\
\begin{subfigure}[h]{0.49\linewidth}
\centering
\includegraphics[width=\linewidth]{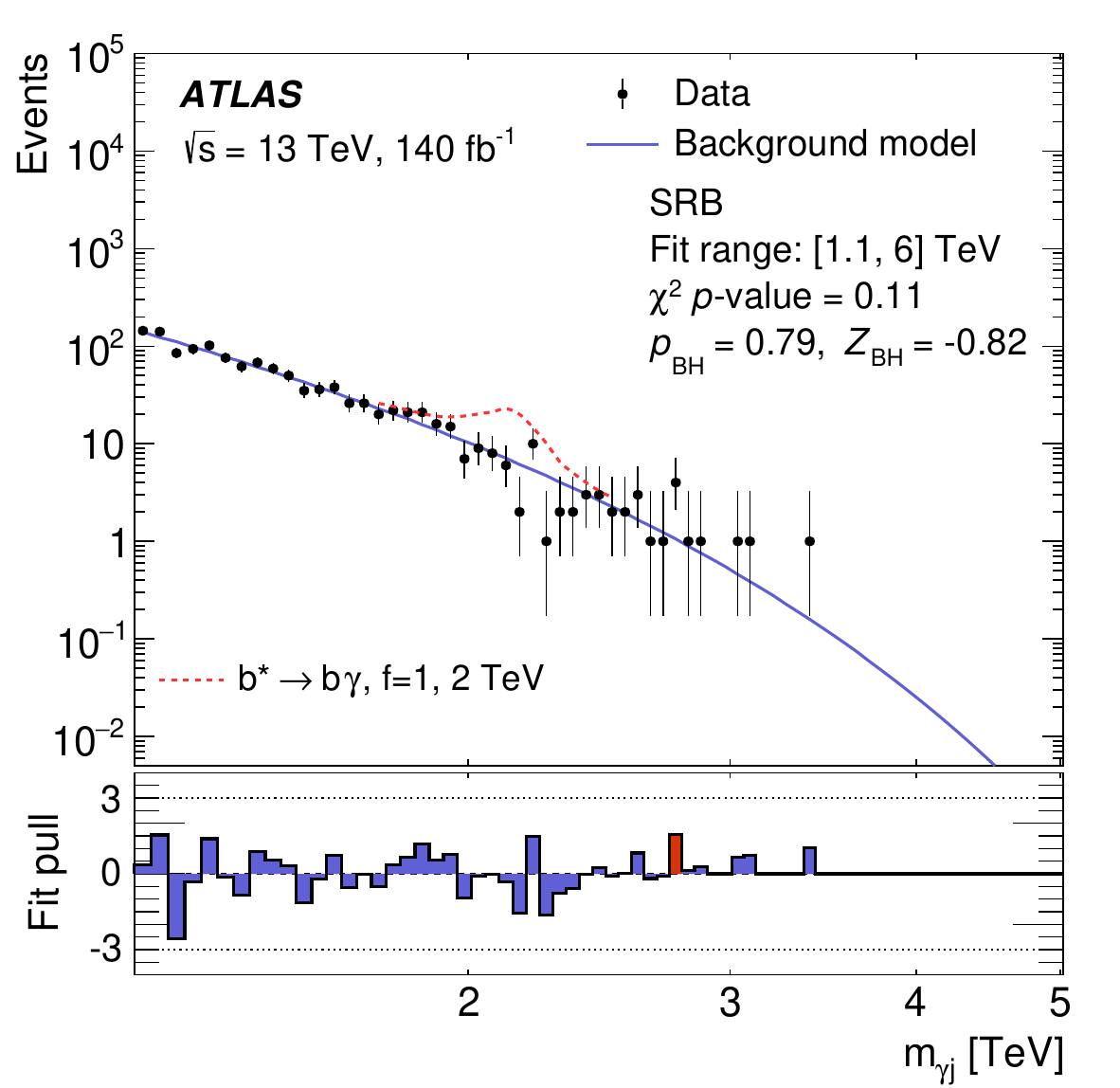}
\caption{}
\end{subfigure}
\begin{subfigure}[h]{0.49\linewidth}
\centering
\includegraphics[width=\linewidth]{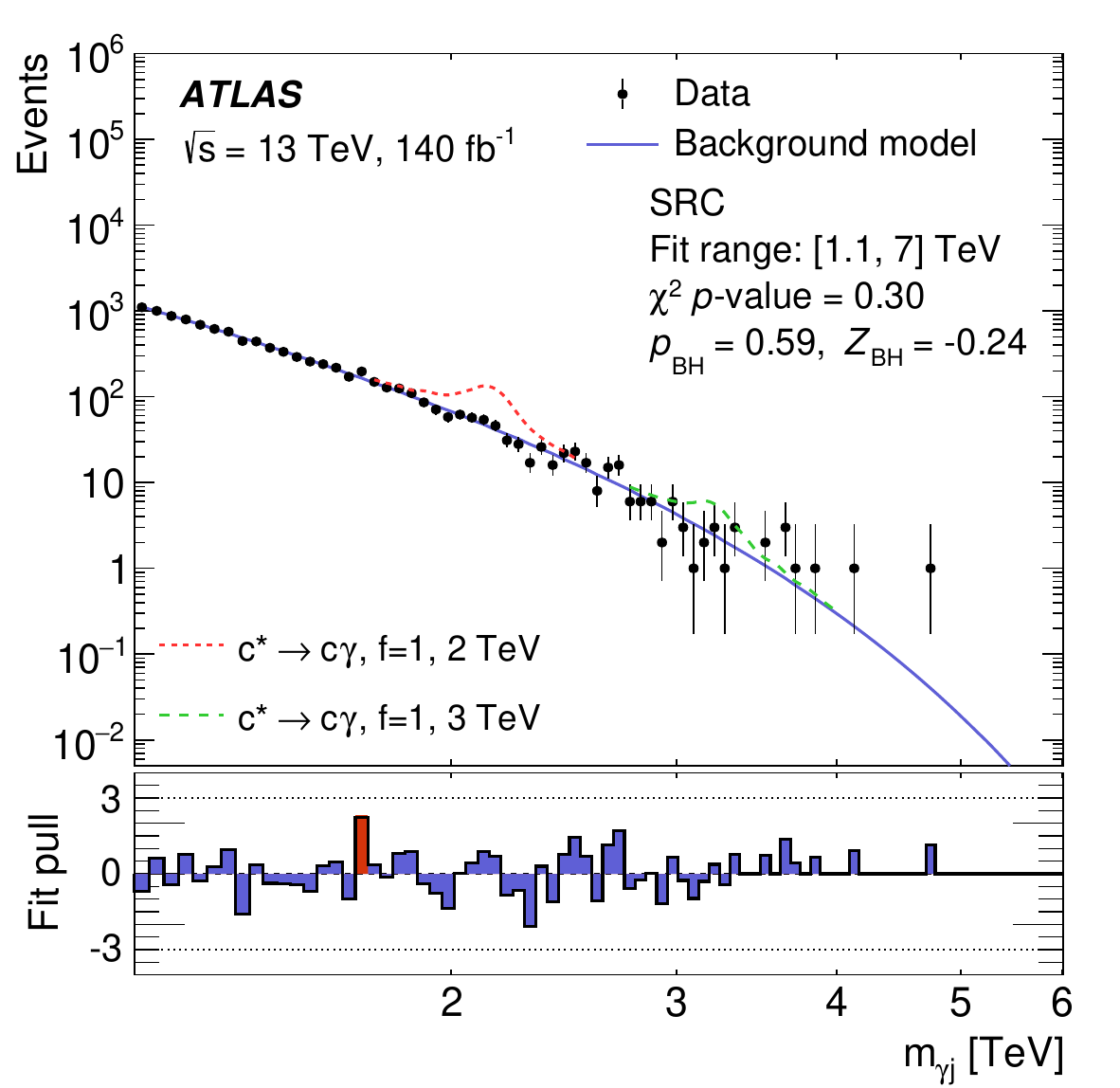}
\caption{}
\end{subfigure}
\caption{Background-only fits to full Run-2 data in the four analysis regions. The data distribution and the background fit are shown in the top panel. Additionally some predicted signal distributions (dashed lines) for excited quarks and quantum black holes are shown in the corresponding region as reference.
The most significant localised excesses identified by the \bh algorithm are highlighted with filled orange bins in the lower panels in the normalised residuals (or pull). Moreover, the $\chi^{2}$ \pval of the fit is displayed for each region along with the \bh \pval and significance.
}
\label{fig:results:bkg_only:bkg_only_fits}
\end{figure}

\subsection{Exclusion Limits}

Since no significant signal is observed, upper limits are set on the production of excited quark and QBH signals decaying into a photon, and a quark or gluon. The expected and observed \(95\%\) confidence level (CL) exclusion limits on the product of the cross-section times branching ratio are derived using a modified frequentist CL$_\text{s}$ method~\cite{Read:2002hq}. Additionally, limits are set on the visible cross-section ($\sigma \times \mathcal{B} \times A \times \varepsilon $) using generic Gaussian-shaped resonances, offering a model-independent approach to interpreting potential new physics signals. This method ensures that the results remain broadly applicable, allowing constraints to be placed on a wide range of theoretical scenarios beyond the specific models studied.
\Tab{\ref{tab:results_limits}} summarises the observed and expected limits for the excited quark and QBH signal models considered, along with the signal regions used in the exclusion fits.

\subsubsection{Excited quarks}

Excited quark signals are separated into three distinct flavours: \lstar, \cstar, and \bstar.
The limits are computed using the asymptotic approximation up to masses of 5 TeV as the results obtained from pseudo-experiments are found to be consistent. For masses above 5 TeV, where no events are observed in the data, the asymptotic approximation is not reliable. In this regime, the limits are therefore
computed using pseudo-experiments ensuring a correct treatment.
For \lstar signals, upper limits are obtained in the inclusive SR region and computed for different coupling values. The expected and observed limits are presented for coupling $f=1$ in \Fig{\ref{fig:results:sigbkg:qstar:limits:SR}}. Local excesses are observed at lower masses (\(\sim1800~\gev\) and \(\sim2600~\gev\)), though they remain within the \(2\sigma\) uncertainty of the expected limit. These excesses coincide with the most significant bumps identified by the \bh algorithm.

\begin{figure}[ht!]
\centering
\begin{subfigure}[t]{0.49\linewidth}
\centering
\includegraphics[width=\linewidth]{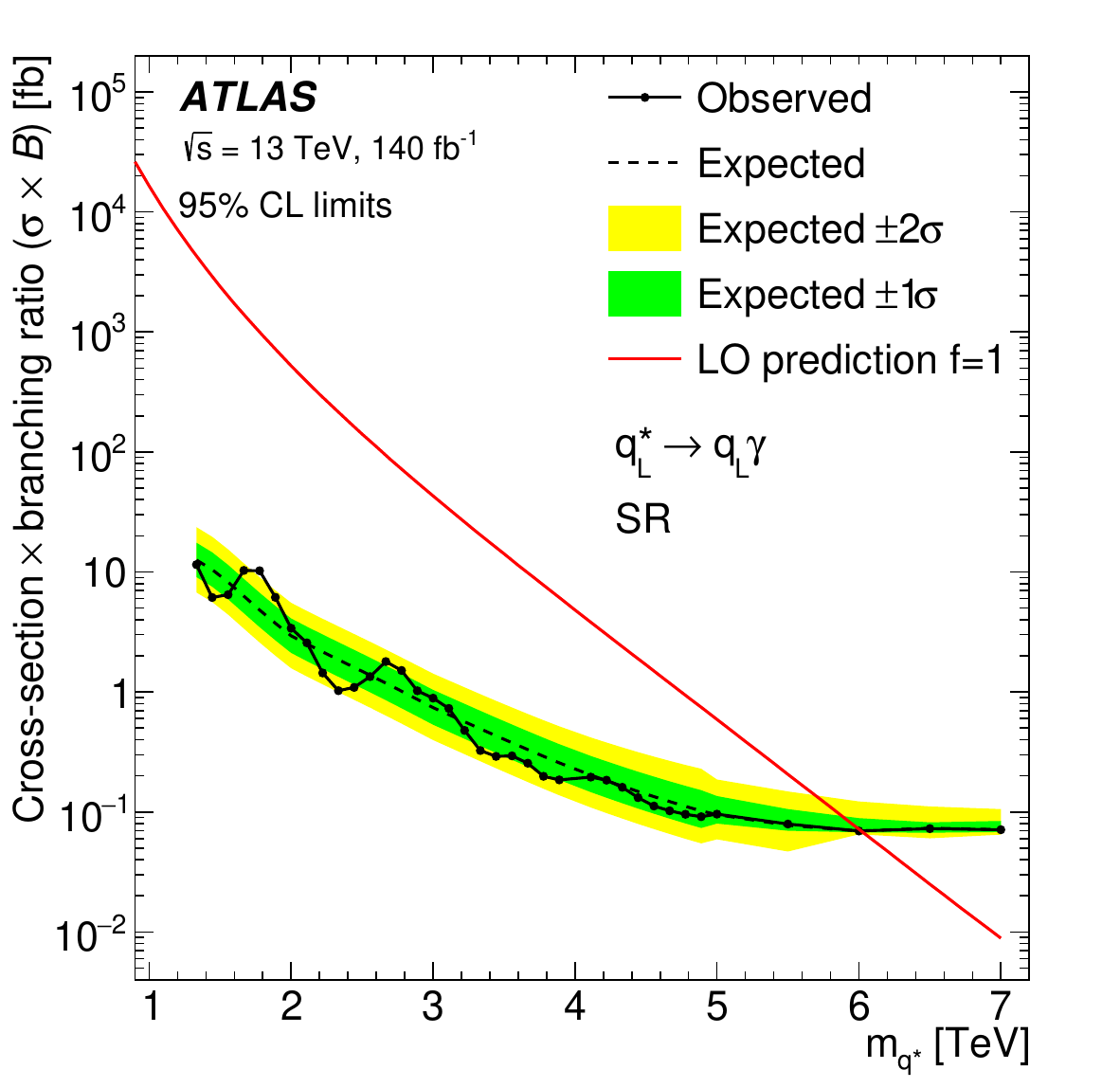}
\caption{}
\label{fig:results:sigbkg:qstar:limits:SR}
\end{subfigure}
\hfill
\begin{subfigure}[t]{0.49\linewidth}
\centering
\includegraphics[width=\linewidth]{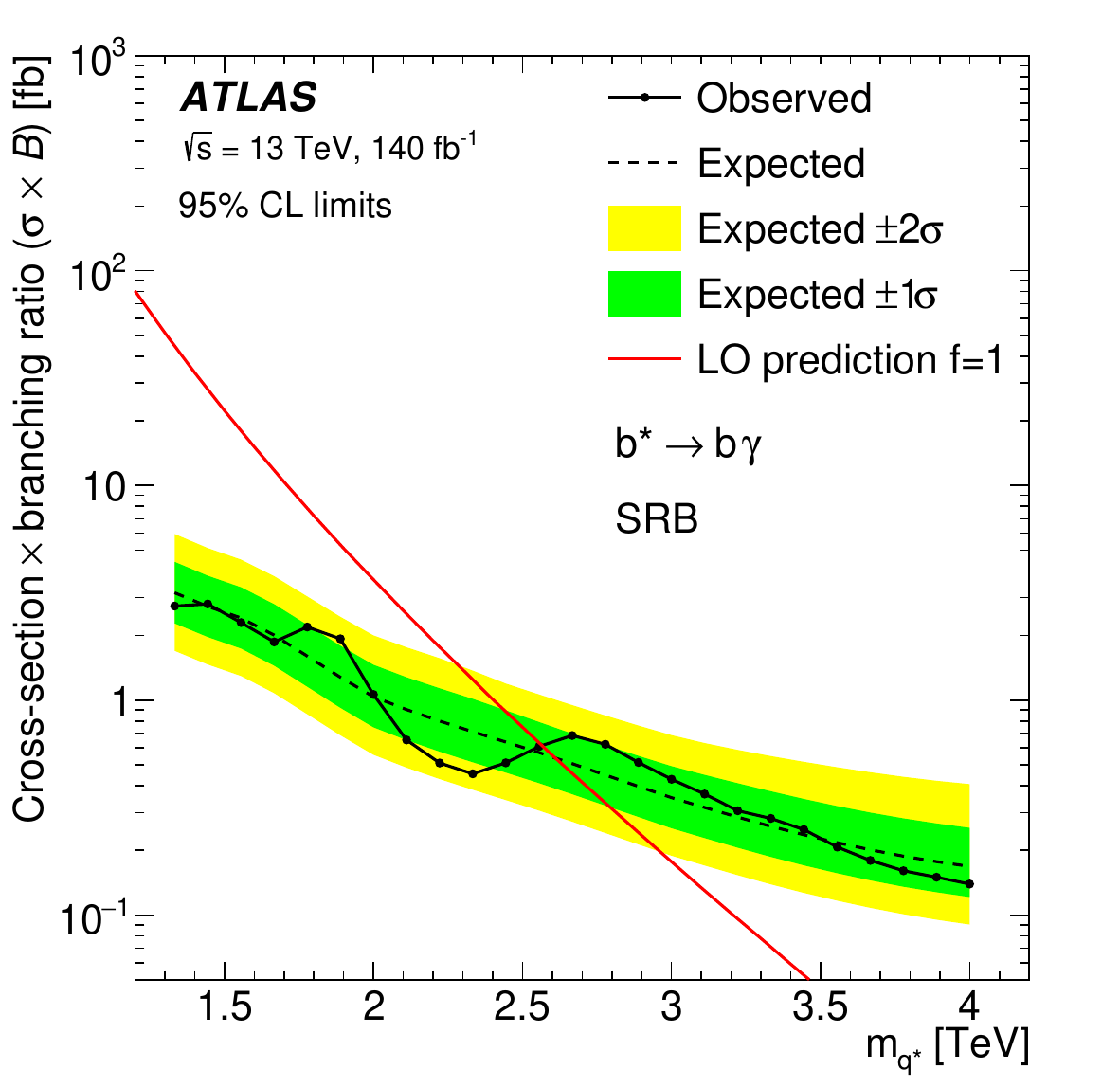}
\caption{}
\label{fig:results:sigbkg:qstar:limits:SRB}
\end{subfigure}\\
\begin{subfigure}[t]{0.49\linewidth}
\centering
\includegraphics[width=\linewidth]{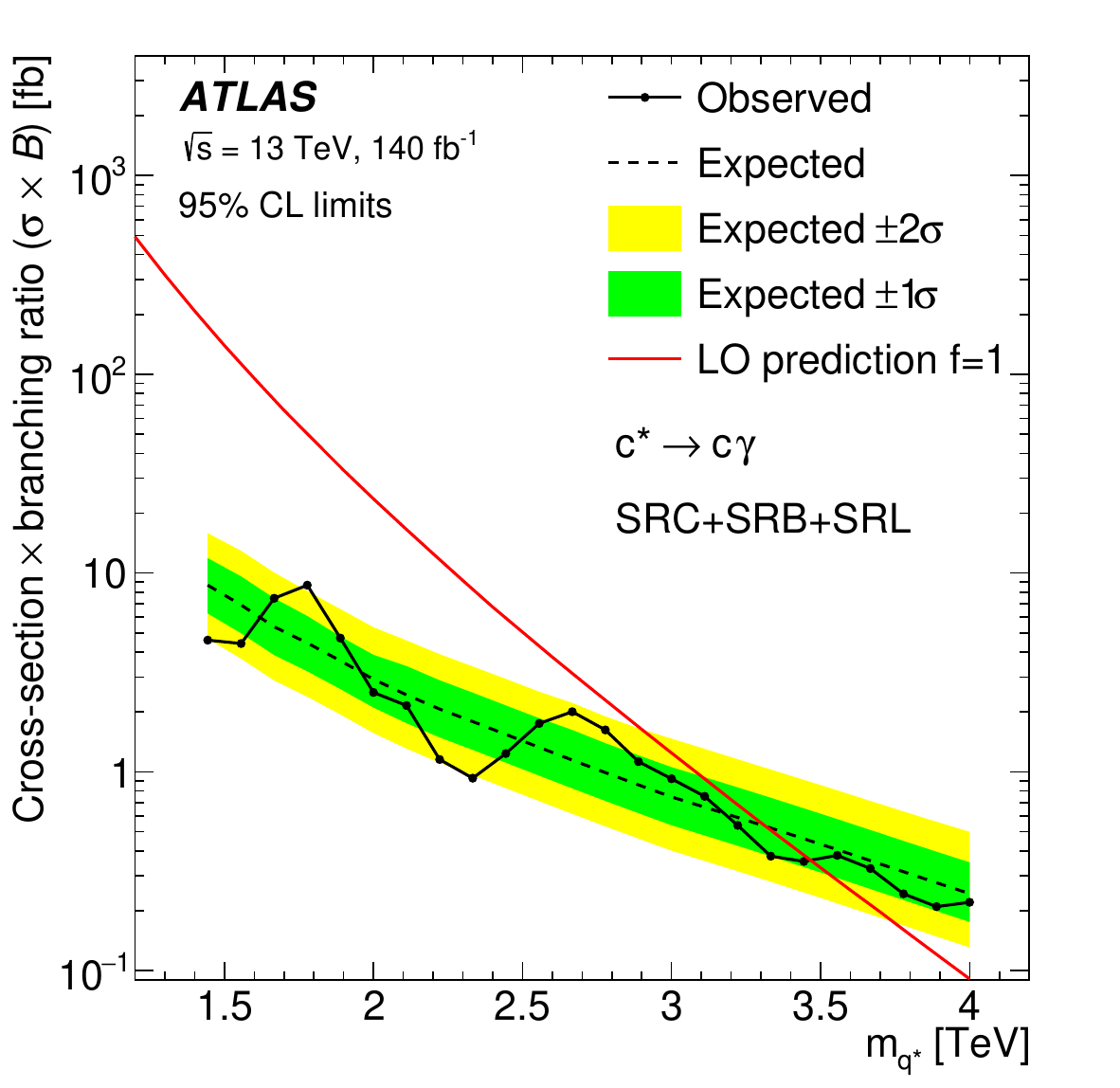}
\caption{}
\label{fig:results:sigbkg:qstar:limits:SRC50}
\end{subfigure}
\caption{Observed and expected upper limits on the cross-section times branching ratio ($\sigma \times \mathcal{B}$)
for (a) \lstar, (b) \bstar and (c) \cstar signals with coupling \(f=1\), using the full Run-2 dataset. Expected upper limits are shown with the black dashed lines, while the line with marks represent observed limits. Red solid lines indicate the predicted cross section for $f=1$.}
\end{figure}

For \bstar signals, the dedicated SRB region is used. The corresponding upper limits are shown in \Fig{\ref{fig:results:sigbkg:qstar:limits:SRB}}, exhibiting a similar pattern to the SR region, where the most significant deviations from the expected limits align with the locations of the most prominent bumps identified by \bh.

This study provides the first limits on \cstar signals in the \yj final state. The upper limits for \cstar are derived using a simultaneous fit across three orthogonal regions: SRC, SRB, and SRL, with results for coupling \(f=1\) illustrated in \Fig{\ref{fig:results:sigbkg:qstar:limits:SRC50}}.

For the benchmark signal excited quarks models considered in this analysis, \lstar, \cstar and \bstar, the presence
of a signal with a mass below \(6.0~\tev\), \(3.5~\tev\) and \(2.6~\tev\), respectively, can be excluded at \(95\%\) CL. These results represent the most stringent limits achieved to date.

It is well understood that the general shape of the signals remains largely unchanged with varying coupling values. Specifically, the core signal structure remains consistent, while higher couplings introduce a secondary off-shell peak at lower \myj, resulting in slightly reduced acceptances for larger \(f\)~\cite{Bhattacharya:2009p2168}.

The results are summarised for all couplings in the excited quark mass-coupling plane in \Fig{\ref{fig:results:sigbkg:qstar:limits_2d}}. This visualization offers a comprehensive overview of the sensitivity to excited quark production, illustrating the coverage across different mass values and coupling strengths for the various signal regions.

\begin{figure}[ht!]
\centering
\includegraphics[width=0.7\linewidth]{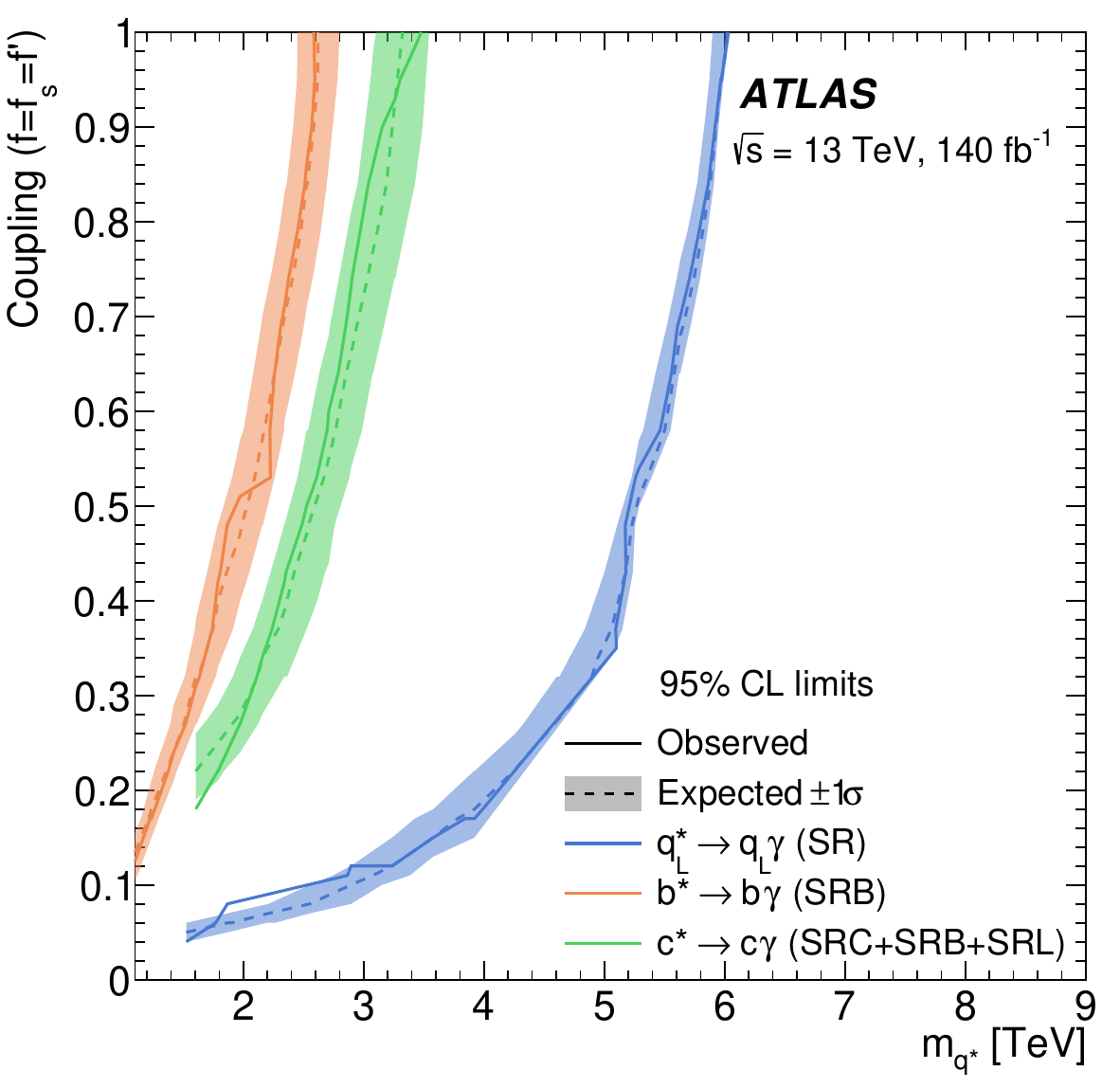}
\caption{Observed and expected upper limits in the mass-coupling plane for the \lstar (SR), \bstar (SRB) and \cstar (SRC+SRB+SRL) signals. The expected (observed) upper limits are displayed with dashed (solid) lines, and the \(1\sigma\) uncertainty band is represented by the coloured shaded regions.}
\label{fig:results:sigbkg:qstar:limits_2d}
\end{figure}

\subsubsection{Quantum Black Holes}

The expected and observed upper limits on the QBH production for both RS1 and ADD models are presented in \Fig{\ref{fig:results:sigbkg:qbh:limits}}. The results are obtained in the inclusive region (SR). Local deviations of up to \(2\sigma\) are observed at {\mth} around \(1.8~\tev\) and \(2.5~\tev\), corresponding to statistically insignificant excesses identified using the \bh algorithm.
For masses below (above) \(4~\tev\), the limits are computed using asymptotic approximation (pseudo-experiments) ensuring a correct treatment.
Signal mass thresholds for RS1 and ADD QBHs below \(5.3~\tev\) and \(7.5~\tev\), respectively, are excluded at \(95\%\) CL. These results represent the most stringent constraints to date on QBH production in the \yj final state.

\begin{figure}[ht!]
\centering
\begin{subfigure}[h]{0.49\linewidth}
\centering
\includegraphics[width=\linewidth]{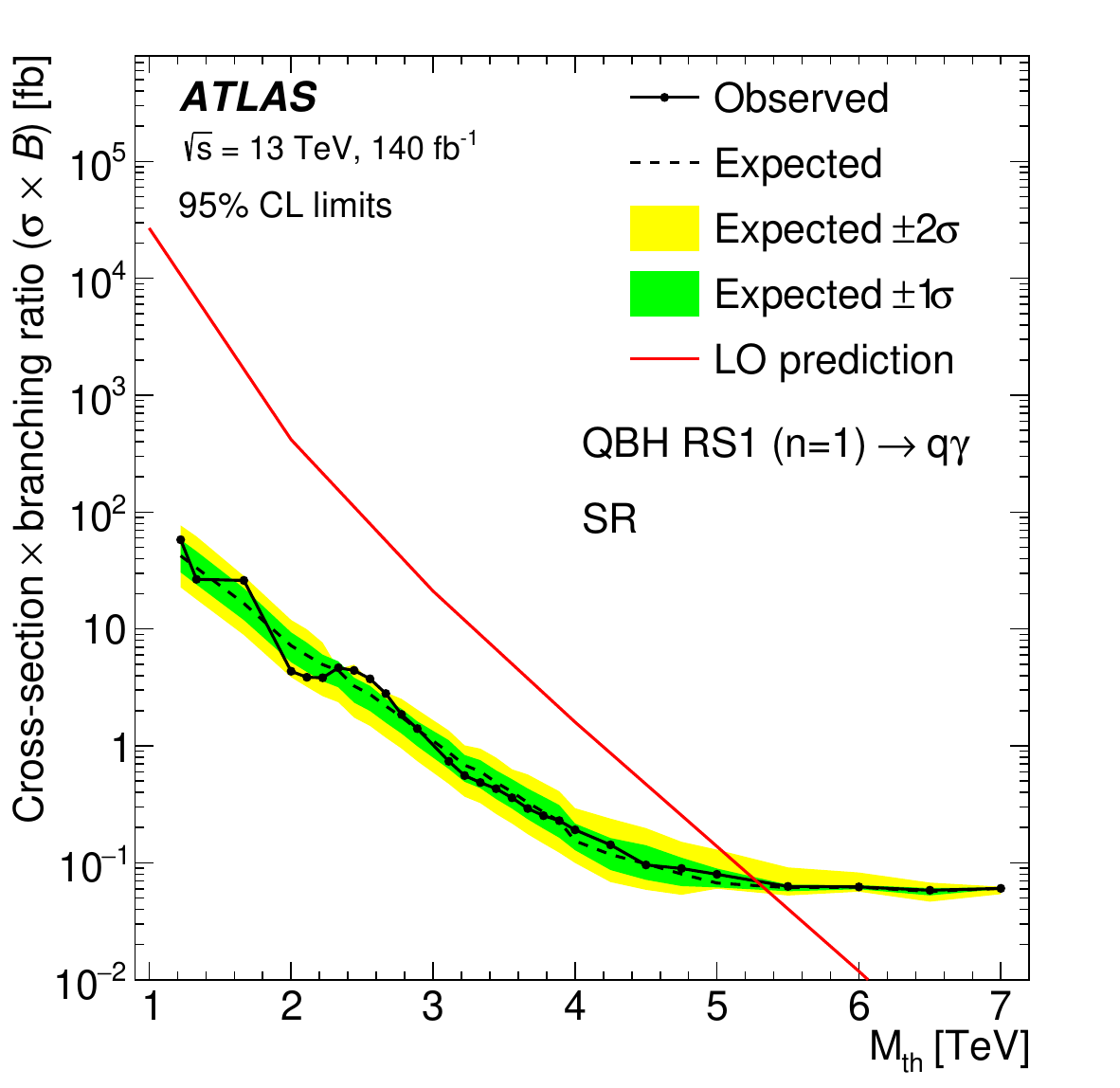}
\caption{}
\end{subfigure}
\hfill
\begin{subfigure}[h]{0.49\linewidth}
\centering
\includegraphics[width=\linewidth]{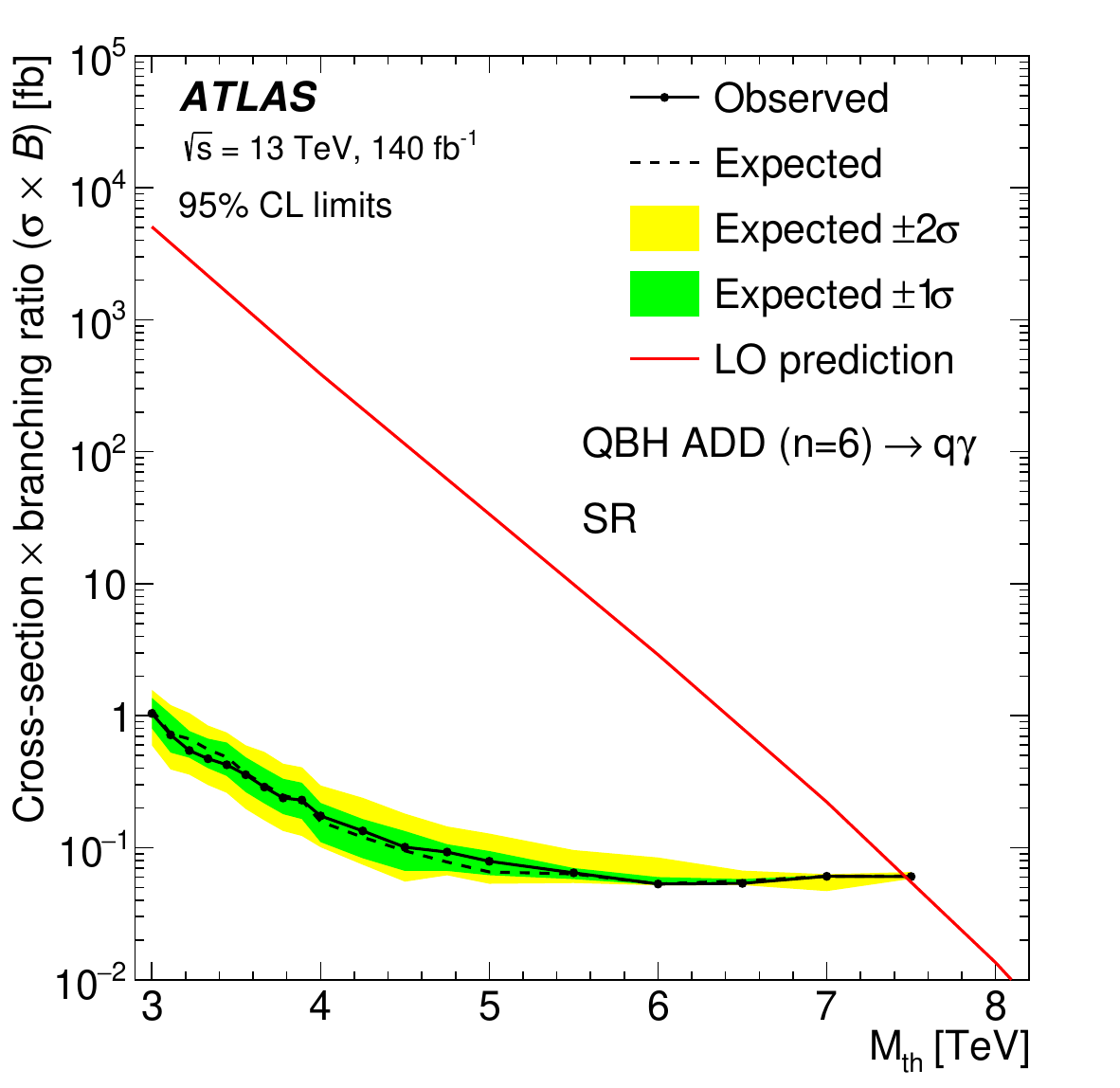}
\caption{}
\end{subfigure}
\caption{Observed and expected upper limits on the cross-section times branching ratio ($\sigma \times \mathcal{B}$) for QBH signal models. The RS1 model (\(n=1\)) is shown on (a), while the ADD model (\(n=6\)) is presented on (b). Observed limits are shown with the black dots and solid line, meanwhile, expected upper limits are represented with the black dashed line, with the corresponding \(\pm 1\) and \(\pm 2 \sigma\) uncertainty bands.}
\label{fig:results:sigbkg:qbh:limits}
\end{figure}

\subsubsection{Generic Gaussian signals}

Generic Gaussian signals enable a model-independent and comprehensive search within this final state. Expected and observed upper limits on the visible cross section for three different Gaussian widths, $\sigma_G/\mG = 0.02,\, 0.07$ and \(0.15\),  analysed across each signal region (SR, SRL, SRC, and SRB), are presented in \Fig{\ref{fig:results:sigbkg:gaus:limits}}. The results in the inclusive SR exhibit a similar structure at low invariant masses around \(1.8~\tev\), and \(2.6~\tev\), as observed in the other benchmark models, with the small excess identified by the \bh algorithm.

\begin{figure}[ht!]
\centering
\begin{subfigure}[h]{0.49\linewidth}
\centering
\includegraphics[width=0.8\linewidth]{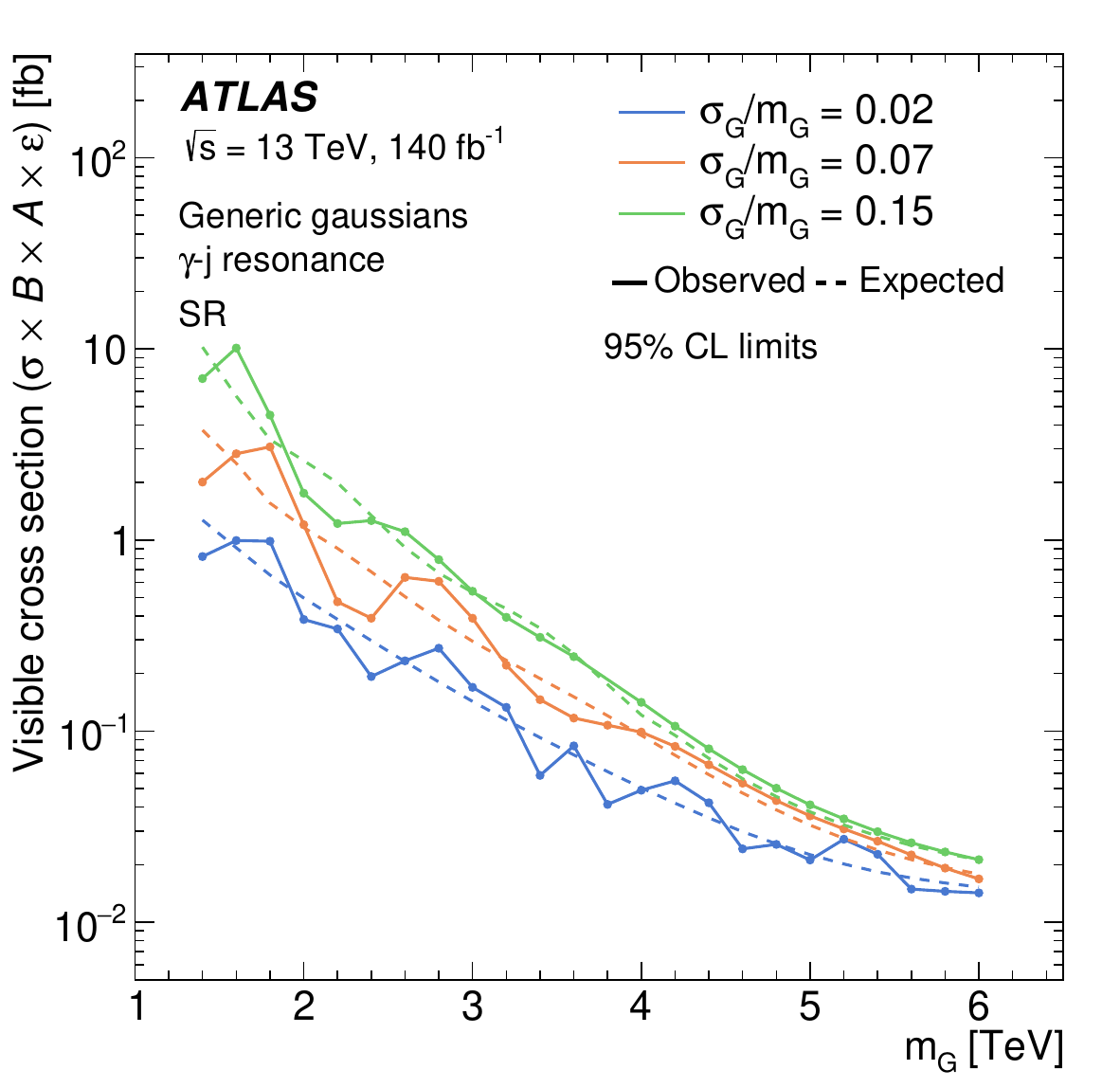}
\caption{}
\end{subfigure}
\begin{subfigure}[h]{0.49\linewidth}
\centering
\includegraphics[width=0.8\linewidth]{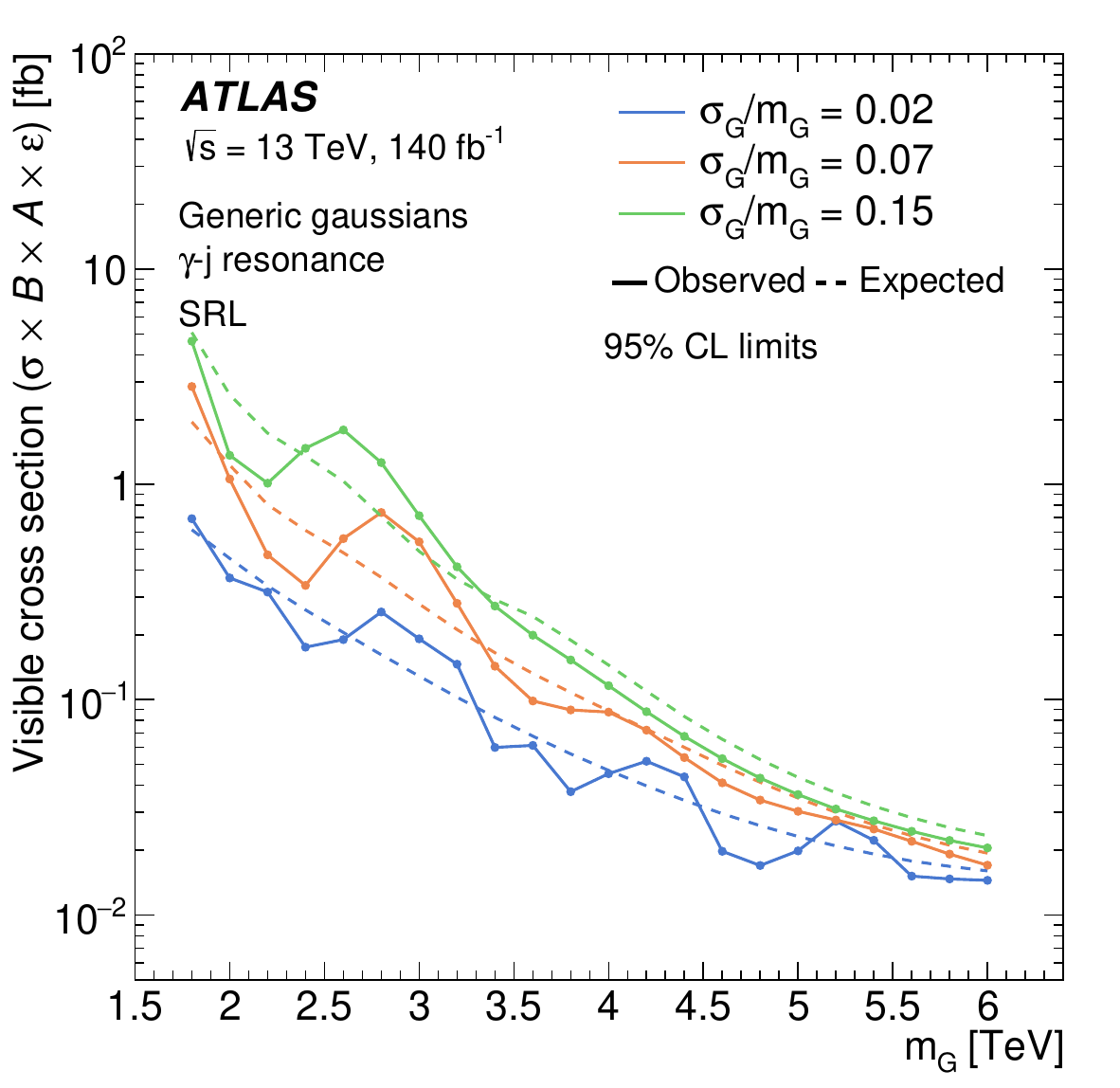}
\caption{}
\end{subfigure}
\begin{subfigure}[h]{0.49\linewidth}
\centering
\includegraphics[width=0.8\linewidth]{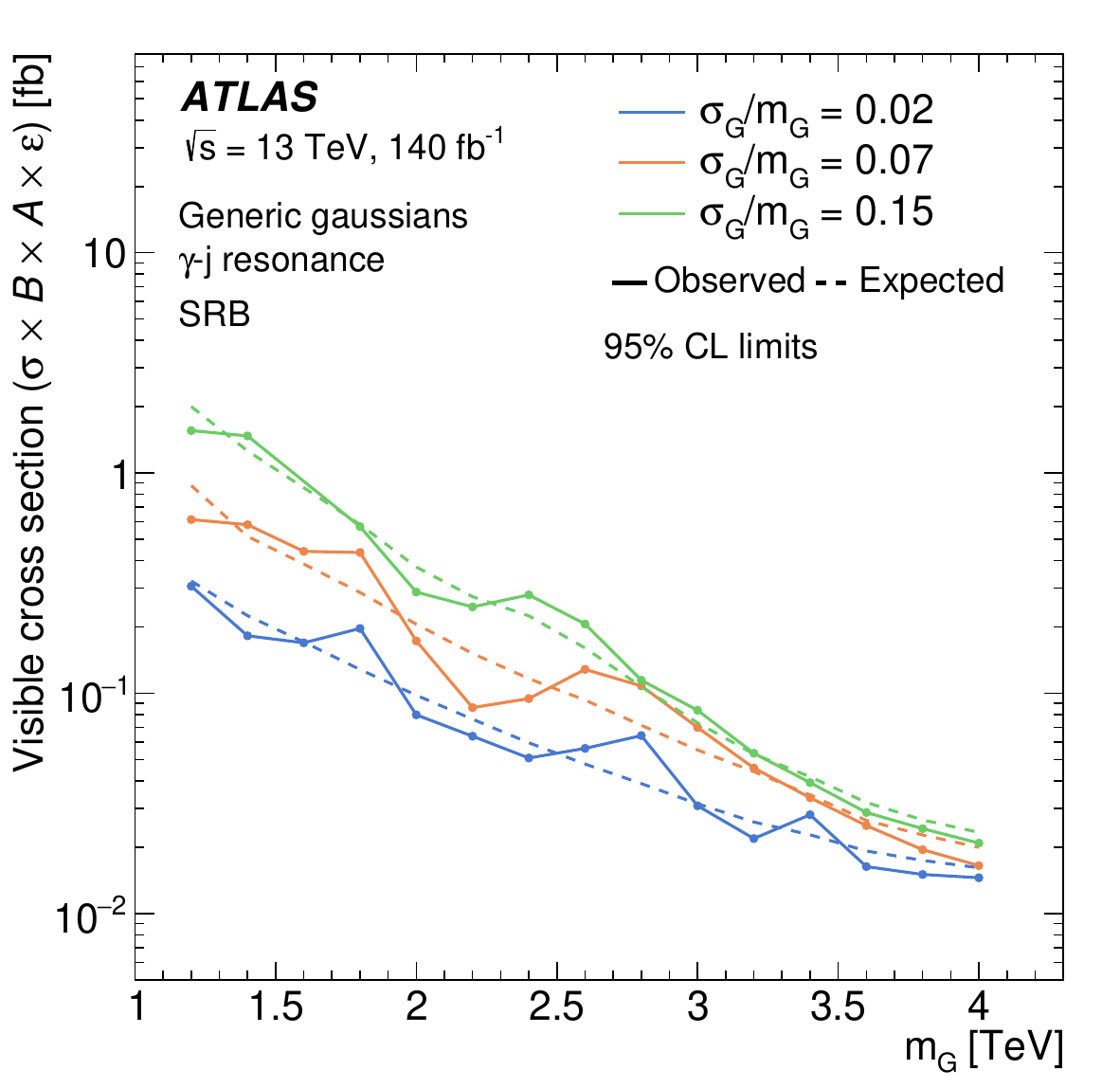}
\caption{}
\end{subfigure}
\begin{subfigure}[h]{0.49\linewidth}
\centering
\includegraphics[width=0.8\linewidth]{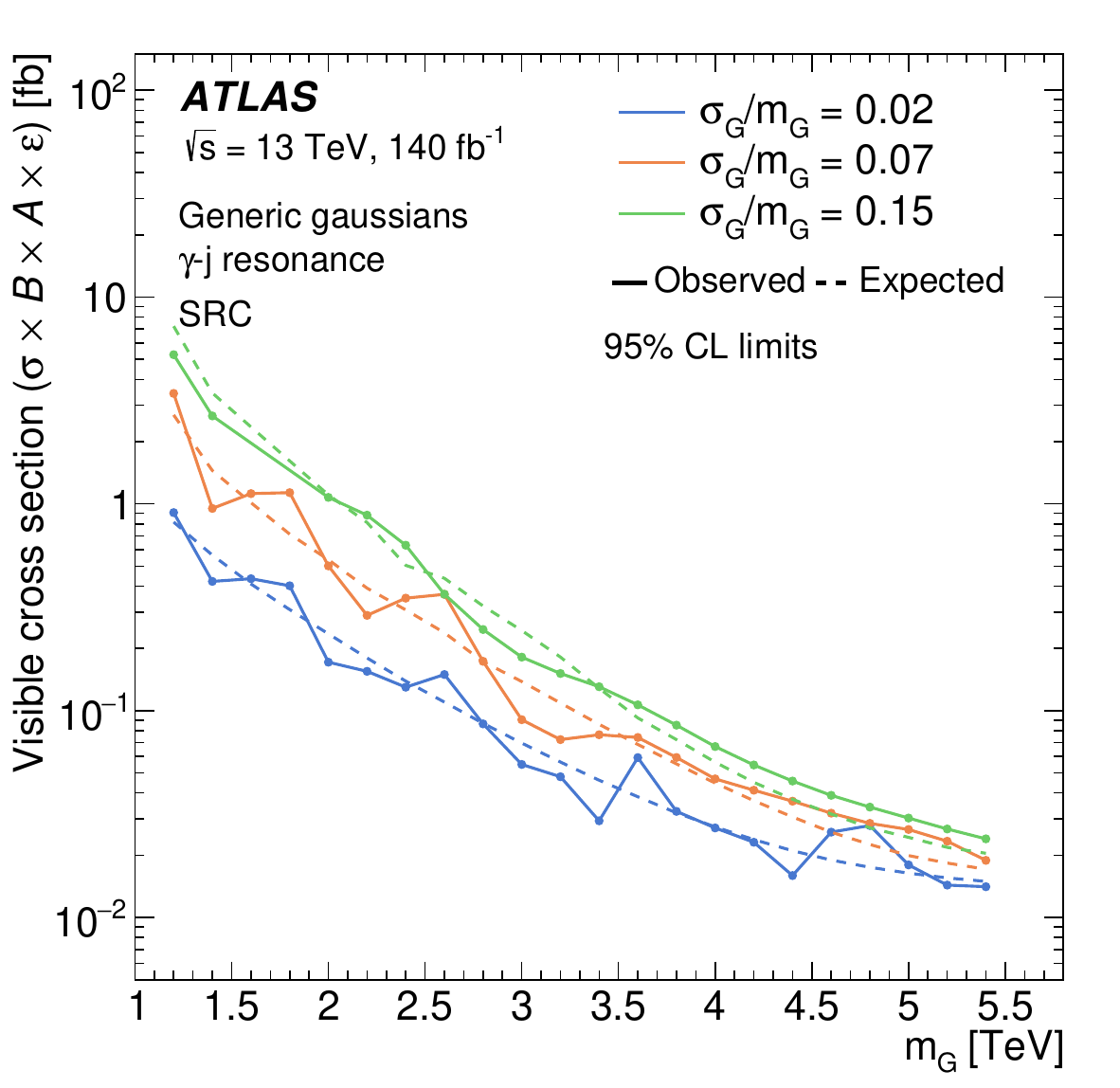}
\caption{}
\end{subfigure}
\caption{Observed and expected upper limits on the visible cross-section ($\sigma \times \mathcal{B} \times A \times \varepsilon$) using generic Gaussian signals with three different widths: $\sigma_G/m_G $ = 0.02, 0.07 and 0.15 in regions (a) SR, (b) SRL, (c) SRB and (d) SRC.}
\label{fig:results:sigbkg:gaus:limits}
\end{figure}

\begin{DIFnomarkup}
\begin{table}
\centering
\caption{Summary of exclusion limits in the different excited quarks and QBH signal models considered. For each signal model the signal regions used in the exclusion fit, the expected and observed limit are shown. All limits at 95\% CL.}
\label{tab:results_limits}
\begin{tabular}{lcrr}
\toprule
\multirow{2}{*}{Signal model}       &      \multirow{2}{*}{Signal Region(s)} & \multicolumn{2}{c}{Mass limit [TeV]} \\
\cmidrule(lr){3-4}
&                       & Expected & Observed \\
\midrule
\(q_{\mathrm{L}}^*, f=1\)       &                SR   & $6.0$  & $6.0$  \\
\(c^*, f=1\)                    &       SRC+SRB+SRL   & $3.3$  & $3.5$  \\
\(b^*, f=1\)                    &               SRB   & $2.6$  & $2.6$  \\
\midrule
QBH RS1 \((n=1)\)               &                 SR  & $5.3$ & $5.3$ \\
QBH ADD \((n=6)\)               &                 SR  & $7.5$ & $7.5$ \\
\bottomrule
\end{tabular}
\end{table}
\end{DIFnomarkup}


\clearpage
\section{Conclusion}
\label{sec:conclusion}


A search for high-mass resonances decaying into a photon and a jet was performed using $140$ {\ifb} of {\pp} collision data at $\sqrt{s} = 13$ {\tev} collected by ATLAS during LHC Run 2.
No significant excess over the Standard Model prediction was observed, and upper limits were set on the production cross sections of new physics scenarios  in both model-dependent and model-independent interpretations.
Excited quarks both light (\(u^{*}/d^{*}\)) and  heavy-flavour, {\cstar} and {\bstar}, were studied, excluding masses up to 6.0 {\tev} ({\lstar}), 3.5 {\tev} ({\cstar}), and 2.6 \tev ({\bstar}), for coupling $f=1$. The {\cstar} limits represent the first constraints on these processes in the LHC experiments. Quantum black hole (Randall-Sundrum  and Arkani-Hamed-Dimopoulos-Dvali) models were also tested, excluding mass thresholds up to 5.3 TeV (RS1) and 7.5 TeV (ADD) at 95\% confidence level. Additionally, upper limits were set on the visible cross-section for generic Gaussian-shaped resonances.
The results presented in this analysis constitute a substantial improvement over previous ATLAS constraints, achieved through the exploitation of the full Run 2 dataset, the application of optimised event selection criteria, and improved background modelling techniques, leading to the strongest constraints to date in the {\yj} final state.


%
%
%
%
%
\section*{Acknowledgements}
%

%

%
%

%
%

We thank CERN for the very successful operation of the LHC and its injectors, as well as the support staff at
CERN and at our institutions worldwide without whom ATLAS could not be operated efficiently.

The crucial computing support from all WLCG partners is acknowledged gratefully, in particular from CERN, the ATLAS Tier-1 facilities at TRIUMF/SFU (Canada), NDGF (Denmark, Norway, Sweden), CC-IN2P3 (France), KIT/GridKA (Germany), INFN-CNAF (Italy), NL-T1 (Netherlands), PIC (Spain), RAL (UK) and BNL (USA), the Tier-2 facilities worldwide and large non-WLCG resource providers. Major contributors of computing resources are listed in Ref.~\cite{ATL-SOFT-PUB-2026-001}.

We gratefully acknowledge the support of ANPCyT, Argentina; YerPhI, Armenia; ARC, Australia; BMWFW and FWF, Austria; ANAS, Azerbaijan; CNPq and FAPESP, Brazil; NSERC, NRC and CFI, Canada; CERN; ANID, Chile; CAS, MOST and NSFC, China; Minciencias, Colombia; MEYS CR, Czech Republic; DNRF and DNSRC, Denmark; IN2P3-CNRS and CEA-DRF/IRFU, France; SRNSFG, Georgia; BMFTR, HGF and MPG, Germany; GSRI, Greece; RGC and Hong Kong SAR, China; ICHEP and Academy of Sciences and Humanities, Israel; INFN, Italy; MEXT and JSPS, Japan; CNRST, Morocco; NWO, Netherlands; RCN, Norway; MNiSW, Poland; FCT, Portugal; MNE/IFA, Romania; MSTDI, Serbia; MSSR, Slovakia; ARIS and MVZI, Slovenia; DSI/NRF, South Africa; MICIU/AEI, Spain; SRC and Wallenberg Foundation, Sweden; SERI, SNSF and Cantons of Bern and Geneva, Switzerland; NSTC, Taipei; TENMAK, T\"urkiye; STFC/UKRI, United Kingdom; DOE and NSF, United States of America.

Individual groups and members have received support from BCKDF, CANARIE, CRC and DRAC, Canada; CERN-CZ, FORTE and PRIMUS, Czech Republic; COST, ERC, ERDF, Horizon 2020, ICSC-NextGenerationEU and Marie Sk{\l}odowska-Curie Actions, European Union; Investissements d'Avenir Labex, Investissements d'Avenir Idex and ANR, France; DFG and AvH Foundation, Germany; Herakleitos, Thales and Aristeia programmes co-financed by EU-ESF and the Greek NSRF, Greece; BSF-NSF and MINERVA, Israel; NCN and NAWA, Poland; La Caixa Banking Foundation, CERCA and AGAUR programs from Generalitat de Catalunya and PROMETEO and GenT Programmes Generalitat Valenciana, Spain; G\"{o}ran Gustafssons Stiftelse, Sweden; The Royal Society and Leverhulme Trust, United Kingdom; Eric and Wendy Schmidt Fund for Strategic Innovation, United States of America.

In addition, individual members wish to acknowledge support from CERN: European Organization for Nuclear Research (CERN DOCT); Chile: Agencia Nacional de Investigaci\'on y Desarrollo (ANID FONDECYT reg. 1230987, FONDECYT 1230812, FONDECYT 1240864, Fondecyt 3240661, Fondecyt Regular 1240721); China: Chinese Ministry of Science and Technology (MOST-2023YFA1605700, MOST-2023YFA1609300), National Natural Science Foundation of China (NSFC - 12175119, NSFC 12275265); Czech Republic: Czech Science Foundation (GACR - 24-11373S), Ministry of Education Youth and Sports (ERC-CZ-LL2327, FORTE CZ.02.01.01/00/22\_008/0004632), PRIMUS Research Programme (PRIMUS/21/SCI/017); EU: H2020 European Research Council (ERC - 101002463); European Union: European Research Council (BARD No. 101116429, ERC - 948254, ERC 101089007), European Regional Development Fund (HE COFUND GA No.101081355, ERDF), Horizon 2020 Framework Programme (MUCCA - CHIST-ERA-19-XAI-00), European Union, Future Artificial Intelligence Research (FAIR-NextGenerationEU PE00000013), Italian Center for High Performance Computing, Big Data and Quantum Computing (ICSC, NextGenerationEU), Marie Sklodowska-Curie Actions (ERC Horizon-2020, GA 956086, GAP-101168829); France: Agence Nationale de la Recherche (ANR-21-CE31-0013, ANR-21-CE31-0022, ANR-22-EDIR-0002, ANR-24-CE31-0504-01); Germany: Baden-Württemberg Stiftung (BW Stiftung-Postdoc Eliteprogramme), Deutsche Forschungsgemeinschaft (DFG - 469666862, DFG - CR 312/5-2); China: Research Grants Council (GRF); Italy: Istituto Nazionale di Fisica Nucleare (ICSC, NextGenerationEU), Ministero dell'Università e della Ricerca (NextGenEU 153D23001490006 M4C2.1.1, NextGenEU I53D23000820006 M4C2.1.1, NextGenEU I53D23001490006 M4C2.1.1, SOE2024\_0000023); Japan: Japan Society for the Promotion of Science (JSPS KAKENHI  JP25H0063, JSPS KAKENHI JP22H01227, JSPS KAKENHI JP22H04944, JSPS KAKENHI JP22KK0227, JSPS KAKENHI JP24K23939, JSPS KAKENHI JP25H00650, JSPS KAKENHI JP25H01291, JSPS KAKENHI JP25K01011, JSPS KAKENHI JP25K01023); Norway: Research Council of Norway (RCN-314472); Poland: Ministry of Science and Higher Education (IDUB AGH, POB8, D4 no 9722), Polish National Science Centre (NCN 2021/42/E/ST2/00350, NCN OPUS 2023/51/B/ST2/02507, NCN OPUS nr 2022/47/B/ST2/03059, NCN UMO-2019/34/E/ST2/00393, UMO-2022/47/O/ST2/00148, UMO-2023/49/B/ST2/04085, UMO-2023/51/B/ST2/00920, UMO-2024/53/N/ST2/00869); Portugal: Foundation for Science and Technology (FCT); Spain: Agència de Gestió d'Ajuts Universitaris i de Recerca. (AGAUR - 2023 BP 00141), Generalitat Valenciana (ASFAE/2022/008), Ministry of Science and Innovation (MCIN \& NextGenEU PCI2022-135018-2, MICIN \& FEDER PID2021-125273NB, RYC2019-028510-I, RYC2020-030254-I, RYC2021-031273-I, RYC2022-038164-I), Ministerio de Ciencia, Innovación y Universidades/Agencia Estatal de Investigaci\'on (EU NextGenerationEU (PRTR-C17.I1), PID2022-142604OB-C22); Sweden: Carl Trygger Foundation (Carl Trygger Foundation CTS 22:2312), Swedish Research Council (Swedish Research Council 2023-04654, VR 2021-03651, VR 2022-03845, VR 2022-04683, VR 2023-03403, VR 2024-05451), Knut and Alice Wallenberg Foundation (KAW 2018.0458, KAW 2022.0358, KAW 2023.0366); Switzerland: Swiss National Science Foundation (SNSF - PCEFP2\_194658); United Kingdom: Leverhulme Trust (Leverhulme Trust RPG-2020-004), Royal Society (NIF-R1-231091); United States of America: U.S. Department of Energy (ECA DE-AC02-76SF00515), Neubauer Family Foundation.

%
%


%
%
%
%
%
%

%
%

%
%
%
\printbibliography
\clearpage
 
\begin{flushleft}
\hypersetup{urlcolor=black}
{\Large The ATLAS Collaboration}

\bigskip

\AtlasOrcid[0000-0002-6665-4934]{G.~Aad}$^\textrm{\scriptsize 104}$,
\AtlasOrcid[0000-0001-7616-1554]{E.~Aakvaag}$^\textrm{\scriptsize 17}$,
\AtlasOrcid[0000-0002-5888-2734]{B.~Abbott}$^\textrm{\scriptsize 122}$,
\AtlasOrcid[0000-0002-0287-5869]{S.~Abdelhameed}$^\textrm{\scriptsize 85b}$,
\AtlasOrcid[0000-0002-1002-1652]{K.~Abeling}$^\textrm{\scriptsize 55}$,
\AtlasOrcid[0000-0001-5763-2760]{N.J.~Abicht}$^\textrm{\scriptsize 49}$,
\AtlasOrcid[0000-0002-8496-9294]{S.H.~Abidi}$^\textrm{\scriptsize 30}$,
\AtlasOrcid[0009-0003-6578-220X]{M.~Aboelela}$^\textrm{\scriptsize 45}$,
\AtlasOrcid[0000-0002-9987-2292]{A.~Aboulhorma}$^\textrm{\scriptsize 36e}$,
\AtlasOrcid[0000-0001-5329-6640]{H.~Abramowicz}$^\textrm{\scriptsize 156}$
% ...
\bigskip
\\

% ...
$^{17}$Department for Physics and Technology, University of Bergen, Bergen; Norway.\\
% ...
$^{30}$Physics Department, Brookhaven National Laboratory, Upton NY; United States of America.\\
% ...
$^{36}$$^{(a)}$Facult\'e des Sciences Ain Chock, Universit\'e Hassan II de Casablanca;$^{(b)}$Facult\'{e} des Sciences, Universit\'{e} Ibn-Tofail, K\'{e}nitra;$^{(c)}$Facult\'e des Sciences Semlalia, Universit\'e Cadi Ayyad, LPHEA-Marrakech;$^{(d)}$LPMR, Facult\'e des Sciences, Universit\'e Mohamed Premier, Oujda;$^{(e)}$Facult\'e des sciences, Universit\'e Mohammed V, Rabat;$^{(f)}$Institute of Applied Physics, Mohammed VI Polytechnic University, Ben Guerir; Morocco.\\
% ...
$^{45}$Physics Department, Southern Methodist University, Dallas TX; United States of America.\\
% ...
$^{49}$Fakult\"{a}t Physik, Technische Universit{\"a}t Dortmund, Dortmund; Germany.\\
% ...
$^{55}$II. Physikalisches Institut, Georg-August-Universit\"{a}t G\"ottingen, G\"ottingen; Germany.\\
% ...
$^{85}$$^{(a)}$Khalifa University of Science and Technology, Abu Dhabi;$^{(b)}$New York University Abu Dhabi, Abu Dhabi;$^{(c)}$United Arab Emirates University, Al Ain;$^{(d)}$University of Sharjah, Sharjah; United Arab Emirates.\\
% ...
$^{104}$CPPM, Aix-Marseille Universit\'e, CNRS/IN2P3, Marseille; France.\\
% ...
$^{122}$Homer L. Dodge Department of Physics and Astronomy, University of Oklahoma, Norman OK; United States of America.\\
% ...
$^{156}$Raymond and Beverly Sackler School of Physics and Astronomy, Tel Aviv University, Tel Aviv; Israel.\\
% ...
$^{a}$ Also at Affiliated with an institute formerly covered by a cooperation agreement with CERN.\\
$^{b}$ Also at An-Najah National University, Nablus; Palestine.\\
$^{c}$ Also at Borough of Manhattan Community College, City University of New York, New York NY; United States of America.\\
$^{d}$ Also at Center for Interdisciplinary Research and Innovation (CIRI-AUTH), Thessaloniki; Greece.\\
$^{e}$ Also at Centre of Physics of the Universities of Minho and Porto (CF-UM-UP); Portugal.\\
$^{f}$ Also at CERN, Geneva; Switzerland.\\
% ...

\end{flushleft}


%

%
%
\clearpage

\end{document}